\documentclass[acmsmall]{acmart}

\usepackage{makecell}
\usepackage{hyperref}
\usepackage{array}
\usepackage{csquotes}
\usepackage{booktabs}
\usepackage[dvipsnames]{xcolor}
\usepackage{tikz}
\usepackage{caption}
\usepackage{xspace}
\usepackage[normalem]{ulem}
\usepackage{multirow}

\makeatletter
\g@addto@macro\@authornotes{\setcounter{footnote}{2}\footnotetext{Co-corresponding authors.}}
\makeatother

\newcommand{\systemName}{\textsc{AdapTest}\xspace}

\newcommand\jw[1]{#1}

\renewcommand{\sout}[1]{}
\AtBeginDocument{%
  }

\setcopyright{cc}
\setcctype{by}
\acmJournal{PACMHCI}
\acmYear{2026} \acmVolume{10} \acmNumber{8} \acmArticle{ISS013}
\acmMonth{11} \acmDOI{10.1145/3859325}

\begin{document}

\title[\systemName: Supporting Designers in Authoring and Evaluating Environmentally Responsive UIs for XR]{\systemName: Supporting Designers in\\ 
Authoring and Evaluating Environmentally Responsive\\ 
User Interfaces for Extended Reality}

% \affiliation{%

\author{Jungwoo Rhee}
\orcid{0009-0005-3832-2835}
\affiliation{%
  \institution{KAIST}
  \department{Department of Industrial Design}
  \city{Daejeon}
  \country{Republic of Korea}
}
\email{jwoorhee@kaist.ac.kr}

\author{Guhn Lee}
\orcid{0009-0002-6688-2280}
\affiliation{%
  \institution{KAIST}
  \department{Department of Industrial Design}
  \city{Daejeon}
  \country{Republic of Korea}
}
\email{leeguhn@kaist.ac.kr}

\author{Yumin Jung}
\orcid{0009-0005-5230-9290}
\affiliation{%
  \institution{KAIST}
  \department{School of Computing}
  \city{Daejeon}
  \country{Republic of Korea}
}
\email{jungym887@gmail.com}

\author{Mustafa Doga Dogan}
\orcid{0000-0003-3983-1955}
\affiliation{%
  \institution{Adobe Research}
  \city{Basel}
  \country{Switzerland}
}
\email{doga@adobe.com}

\author{Kongpyung (Justin) Moon}
\authornotemark[2]
\orcid{0000-0001-9691-0129}
\affiliation{%
  \institution{National University of Singapore}
  \department{Department of Design}
  \city{Singapore}
  \country{Singapore}
}
\email{justin.moon@nus.edu.sg}

\author{Andrea Bianchi}
\authornotemark[2]
\orcid{0000-0002-7500-7974}
\affiliation{%
  \institution{KAIST}
  \department{Department of Industrial Design}
  \city{Daejeon}
  \country{Republic of Korea}
}
\email{andrea@kaist.ac.kr}

\renewcommand{\shortauthors}{Rhee et al.}

\begin{abstract}
Extended Reality (XR) headsets increasingly host familiar 2D user interfaces as windows floating in physical environments. However, their usability depends on environmental conditions that designers cannot fully control.
Current design workflows often separate how a variant is designed from how it actually looks in the target environment.
We present \systemName, an in-situ authoring and testing tool that lets designers select interface elements, define when and how they change, compare the resulting variations, and test their activation without coding. Based on prior research and a formative study with 12 interface designers, the system targets three key environmental variables: distance, lighting, and occlusion. 
We evaluated \systemName against a desktop design and in-headset preview workflow in a within-subjects study with 12 designers. Compared with the baseline\sout{on a separate task domain}, designers accounted for a wider range of environmental factors, addressed more overlapping conditions, and repeatedly tested and revised their designs in place; there were no significant differences in workload or usability despite longer sessions. Our work contributes an integrated, designer-centered approach to adapting, comparing, and validating environmentally responsive XR user interfaces.

% [OLD REVISION] 2D interfaces placed in Extended Reality (XR) must remain usable as their physical viewing context changes, yet current design workflows often separate visual variants from the conditions that should activate them. Existing approaches automate how XR interfaces adapt, require adaptive behaviors to be implemented in code, or treat design variants as static artifacts, leaving visual designers without a unified way to author and evaluate adaptive behavior. We present AdapTest, an in-situ authoring and testing tool that lets designers select interface elements, define when and how they change, compare the resulting variations, and test their activation without coding. Based on prior research and a formative study with 12 interface designers, the system targets three key environmental variables: distance, lighting, and occlusion. We evaluated AdapTest against a desktop design and in-headset preview workflow in a within-subjects study with 12 designers. Compared with the baseline, participants accounted for more of these environmental factors, created combined adaptations more frequently, and repeatedly tested and revised their designs in place, while workload and usability showed minimal difference despite longer sessions. Our work contributes an integrated, designer-centered approach to adapting, comparing, and validating environmentally responsive XR user interfaces.
\end{abstract}




\begin{teaserfigure}
  \includegraphics[width=\textwidth]{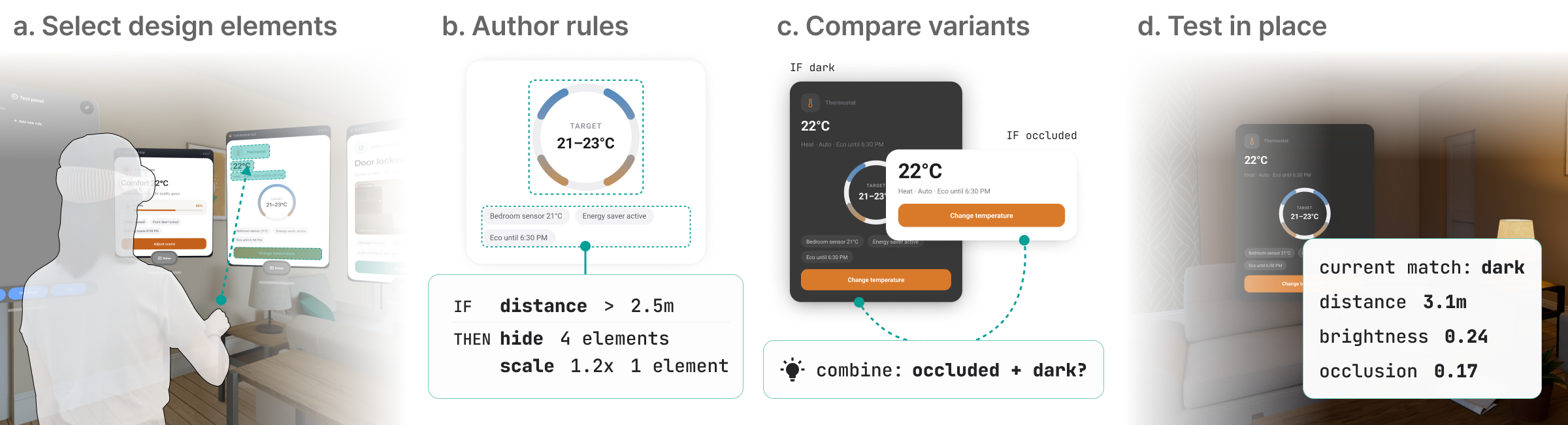}
  \caption{\systemName lets designers make 2D interfaces respond to the physical viewing context in an XR environment. A designer \textbf{(a)} selects an element layer from UI design, \textbf{(b)} authors an IF/THEN rule, \textbf{(c)} previews condition-specific variants attached around the default design, and \textbf{(d)} validates the matched variant in situ as the viewpoint changes.}
  \Description{A four-panel walkthrough. Panel a shows a headset-wearing designer selecting outlined elements from a thermostat interface situated in a room. Panel b shows the selected thermostat elements and a rule that, when distance exceeds 2.5 meters, hides four elements and scales one element by 1.2 times. Panel c compares dark and occluded variants and suggests combining the two conditions. Panel d shows the thermostat in a dark room beside live Test Mode values: the current match is dark, distance is 3.1 meters, brightness is 0.24, and occlusion is 0.17.}
  \label{fig:teaser}
\end{teaserfigure}

\received{25 July 2025}
\received[revised]{24 September 2026}
\received[accepted]{7 October 2026}

\maketitle
\section{Introduction}

Graphical user interfaces are moving off the screen and into physical space.
Extended Reality (XR) headsets now render familiar 2D interfaces---dashboards, control panels, exhibit guides---as windows carried into immersive surroundings \cite{EtherealPlanes, Zhang_2026}, with deployments spanning museums, homes, and workplaces \cite{Khan_2024, UIEntity}.
Unlike a phone or desktop screen, such an interface is at the mercy of its environment: the viewer walks away and body text becomes illegible \cite{Harrison_2008, Vogel_2004}; the room dims and low-contrast styling disappears against passthrough \cite{DarkOSTHMD_2019, DarkOSTHMD_2021}; furniture or passersby occlude the panel \cite{Bell_2001, situationAdapt}; cluttered backdrops erode the readability of overlaid content \cite{2DvizAR, SituatedPatterns}.
Most interface design is fixed at authoring time, so designers have to anticipate these conditions as separate cases. 

Current tools give visual designers partial control over how their interfaces adapt \jw{to the XR environment} and limited means to test that behavior in place.
Automated systems decide the adaptation at runtime---repositioning or anchoring windows to spatial context \cite{AUIT, situationAdapt, davari2024contextawareadaptationextendedreality}, matching layouts to the surrounding scene \cite{semanticAdapt}, varying information density \cite{ContextAdapt}, projecting onto nearby surfaces \cite{blendMR, AdapTUI}, switching input techniques \cite{Adaptique_2025}, or recoloring panels against their background \cite{Santos_2025}.
Because these decisions live in code and optimization objectives, designers cannot see or shape what the interface becomes under each condition, and often cannot picture it in advance \cite{XRProfessionalInterview, Chen_2024}.
Design practice keeps that control in the designer's hands but loses the tie to condition: designers hand-build each case as a duplicated static artboard (e.g., a dark-room, far-away, or occluded version \cite{XRProfessionalInterview, KraussVRST2021, Ashtari_2020}) and preview the copies in headset prototyping tools such as ShapesXR \cite{ShapesXR}.
These copies let a designer inspect a mockup, yet they do not connect a condition to the change it should trigger, and confirming that one behaves correctly in place would mean coding, rebuilding, and redeploying to the headset---development work outside design practice \cite{Ashtari_2020, CoCreatAR}.
Designers are therefore left either to accept automatic adaptation or approximate it by hand, without a way to author their own condition-to-change mappings \jw{for the environmental conditions of a 2D interface in XR} and test them in the room.

To address this gap, we present \systemName, an authoring tool that lets designers adapt the GUIs they already create to the environments those interfaces will inhabit.
After opening a design in the headset, a designer places its screens around the room, examines them where they will be used, and decides when the interface should change as spatial conditions shift.
To author that change, the designer selects an element, assigns an alternative style such as larger type, higher contrast, or fewer details, and defines when that style should apply: when the viewer is far, the room is dark, or the panel is blocked.
Each rule's outcome appears in a preview beside the default design, so the designer compares how the interface should adapt across conditions.
The designer then tests the rules against the room's actual conditions, observes each transition as it fires, and refines the thresholds and styling accordingly.
In a within-subjects study, twelve designers created adaptive versions of multi-screen interfaces with \systemName and with a current-practice baseline pairing a 2D design tool with in-headset preview.
With \systemName, all twelve participants authored explicit conditional rules, created design variants covering multiple environmental factors, and interleaved authoring with repeated Test Mode runs. Workload and usability remained comparable to the existing workflow.
Participants noted that current tools represent adaptive behavior only as text beside the design, whereas \systemName allowed them to author and experience that behavior as a working rule and reuse it across screens.

In summary, this paper contributes (1) a design characterization of environmental rule authoring for GUIs in XR---which elements designers target, which conditions trigger change, which style and layout outputs they reach for, and how far their \sout{scenario coverage extends}\jw{scenarios cover} under live spatial conditions; (2) \systemName, an authoring tool that unifies element styling, environmental rules, \sout{attached}condition previews, and situated testing in a single workflow, so that the \sout{adaptation of 2D interface elements}\jw{2D UI adaptation} stays under designer control, indexed to sensed environment\sout{al conditions}, and testable in place; and (3) a \sout{comparative}\jw{user} study comparing designers' condition coverage, target selection, combination reasoning, and environmental debugging between \systemName and a Figma-based current-practice workflow.

\section{Related Works}

\systemName builds on three research areas: how adaptive behavior for XR interfaces is authored, how designers compare design variants, and how XR designs are tested in place.

\subsection{Authoring Adaptive Behavior for XR Interfaces}

Research on XR adaptivity has made strong progress on automated adaptation, where the system decides at runtime how the interface changes.
AUIT \cite{AUIT} lets XR creators attach objectives, such as visibility and reachability, to UI elements in Unity, and a solver adjusts the \sout{elements' }positions, rotations, and scales at runtime.
SemanticAdapt \cite{semanticAdapt} retargets a world-anchored layout to a new room by matching virtual \sout{elements}to semantically related physical objects, and SituationAdapt \cite{situationAdapt} re-places elements after a vision-language model rates how socially appropriate each surrounding area is.
Further systems select placement strategies by context \cite{davari2024contextawareadaptationextendedreality}, compute layouts against the physical scene \cite{FLARE}, project interfaces onto surrounding geometry \cite{blendMR, AdapTUI}, and switch selection techniques as targets and posture change \cite{Adaptique_2025}.
Automated adaptation has also reached visual properties: correcting color and contrast against the background \cite{SmartColor, BackgroundAwareColor}, repositioning and recoloring panels from sensed scene context \cite{Santos_2025}, adjusting level of detail with the viewer's visual angle \cite{LODAR}, and adapting rendering to the sensed environment in web-based AR \cite{Ferrao_2023}.
Purpose-built interfaces further demonstrate the value of environmental conditions, such as a museum guide that swaps content based on visitor distance \cite{DistMuseum}.
Even when content creators pre-design the variants (e.g., \cite{ContextAdapt}), where applications carry designed level-of-detail versions and an optimizer selects among them by cognitive load, task, and environment, the runtime still owns the mapping from context to variant.
Optimization itself has grown more open to \sout{its }users, who can choose among Pareto-optimal proposals \cite{ParetoAdapt_2023} or steer the objectives by spoken preference \cite{AutoOpt_2026}.
Across these systems, the condition--appearance mapping lives in cost functions, objective priorities, and code, which makes it difficult for visual designers to inspect what will happen under a given condition or to adjust it without programming \cite{XRProfessionalInterview, Chen_2024}.

Rule-based authoring, in contrast, puts the mapping in the author's hands, and XR tools have primarily applied it to interaction events.
Trigger-action programming has long enabled non-programmers to state IF/THEN behavior \cite{IFTTT}.
In XR, XRSpotlight \cite{XRSpotlight} presents the interactions of a Unity scene as natural-language rules that novice developers can inspect and copy across scenes; ProInterAR \cite{ProInterAR} lets creators stack blocks that react to grasps, gestures, and collisions between AR contents; ProGesAR \cite{ProGesAR} lets designers place proxemic and gestural triggers that fire when a person crosses a distance threshold or performs a gesture; and RealityCanvas \cite{RealityCanvas} binds sketched animations to tracked physical motion.
FrameKit \cite{FrameKit_2024} brings authored adaptation to 2D interfaces by placing keyframes along parameters such as viewer distance, with in-between states interpolated and rendered from simulated sensor values.
These works establish that authors can express rich behavior through rules, with triggers as discrete events (e.g., input actions, collisions, threshold crossings) and actions on scene objects or sketches.
Rules that respond to sensed environmental conditions and attach to the \sout{elements of a GUI displayed}\jw{GUI elements} in XR remain an open opportunity.

\systemName extends rule-based authoring \sout{in this direction}\jw{by}: designers select granular UI elements and state how those elements should change under continuous environmental conditions, keeping the condition--variant mapping explicit, deterministic, and editable without code.

\subsection{Comparing Design Variants}

Seeing design alternatives side by side is an established aid to design judgment.
Creating and comparing prototypes in parallel yields better outcomes and more divergent exploration than refining one at a time \cite{Dow_2011}, and tools have operationalized this insight: Juxtapose \cite{Hartmann_2008} runs alternative interface implementations side by side with runtime tuning, subjunctive interfaces let users set up and view multiple \sout{alternative }scenarios in parallel \cite{Lunzer_2008}, Scout \cite{Swearngin_2020} generates layout \sout{alternatives}\jw{variants} from high-level constraints for rapid comparison, and DesignTrace \cite{DesignTrace_2026} tracks alternatives as designers iterate with a generative model.
These tools compare alternatives that the designer freely explores; the variants answer "which design do I prefer?"
Adaptive XR interfaces pose a different comparison question---"does my design hold up under every condition it will meet?"---which current practice can answer only partially.
Designers encode environmental cases as duplicated artboards or component variants \cite{KraussVRST2021, XRProfessionalInterview}, but the mapping from condition to variant remains only in file names and memory.
Responsive web design offers a designer-facing analogy: media queries bind layout variants to viewport conditions, and design tools preview breakpoints side by side.
The same side-by-side preview for the environmental conditions of an XR space is the open question.

To answer this, \systemName produces a condition-specific preview that wraps around the default design, so a designer can inspect coverage across dark-to-bright, far-to-near, occluded-to-visible, and combined conditions at a glance.

\subsection{Testing XR Designs in Place}

Validating an XR design still requires being in the deployment environment, which \sout{makes iteration slow}\jw{slows iteration}.
Studies of XR practitioners report that creators cannot judge legibility, scale, or lighting from the desktop \cite{Ashtari_2020, XRProfessionalInterview}. 
Site-specific work demands "frequent on-site visits and additional iterations" \sout{---in CoCreatAR's formative study }\cite{CoCreatAR}, developers spent over a third of their time testing in situ and identified lighting as one condition that desktop views cannot reproduce.
In-situ authoring tools shorten this loop by moving creation into the target environment: DART \cite{DART} moved AR authoring into the designer's workflow, PintAR \cite{PintAR} sketches interfaces directly in AR, Rapido \cite{Rapido} turns video captures into interactive AR prototypes, ImaginateAR \cite{ImaginateAR_2025} assembles outdoor scenes in place with AI assistance, and commercial tools such as ShapesXR \cite{ShapesXR} preview screen designs spatially in the headset; CAVE-AR \cite{Cavallo_2019} approaches the same challenge from the opposite direction, simulating and debugging AR experiences from within VR.
Fast-Forward Reality \cite{FastForward_2024} generates test cases for context-aware policies and plays as immersive visualizations to verify in XR before deployment.
In these tools---as in the live try-out loops of the rule-based authoring tools above---testing is limited to the conditions the room offers at that moment; a darkened room or a twenty-meter viewing distance is difficult to produce on demand, and the tester has limited means to tell which authored state is currently active.

\systemName's \sout{Test Mode }builds on these testing practices: the designer dials environmental conditions with manual controls or drives them from live headset sensing, and the system activates the currently matched rule---turning situated testing from re-enacting circumstances into querying the rule set.

\section{Formative Study}

We conducted a formative study to observe how screen-rooted designers adapt their interfaces when the designs enter a real space\sout{, before committing to a tool design}.
Twelve designers prototyped interfaces for a multi-zone makerspace and reviewed them as virtual screens on Apple Vision Pro from three viewing distances, iterating in place.
Their behavior yielded three findings---an element-level adaptation vocabulary, a struggle to cover compounding conditions, and a dependence on being in place to judge thresholds and transitions---from which we derive design goals DG1--DG3.

\subsection{Method}

\begin{figure}[h!]
    \centering
    \includegraphics[width=0.9\linewidth]{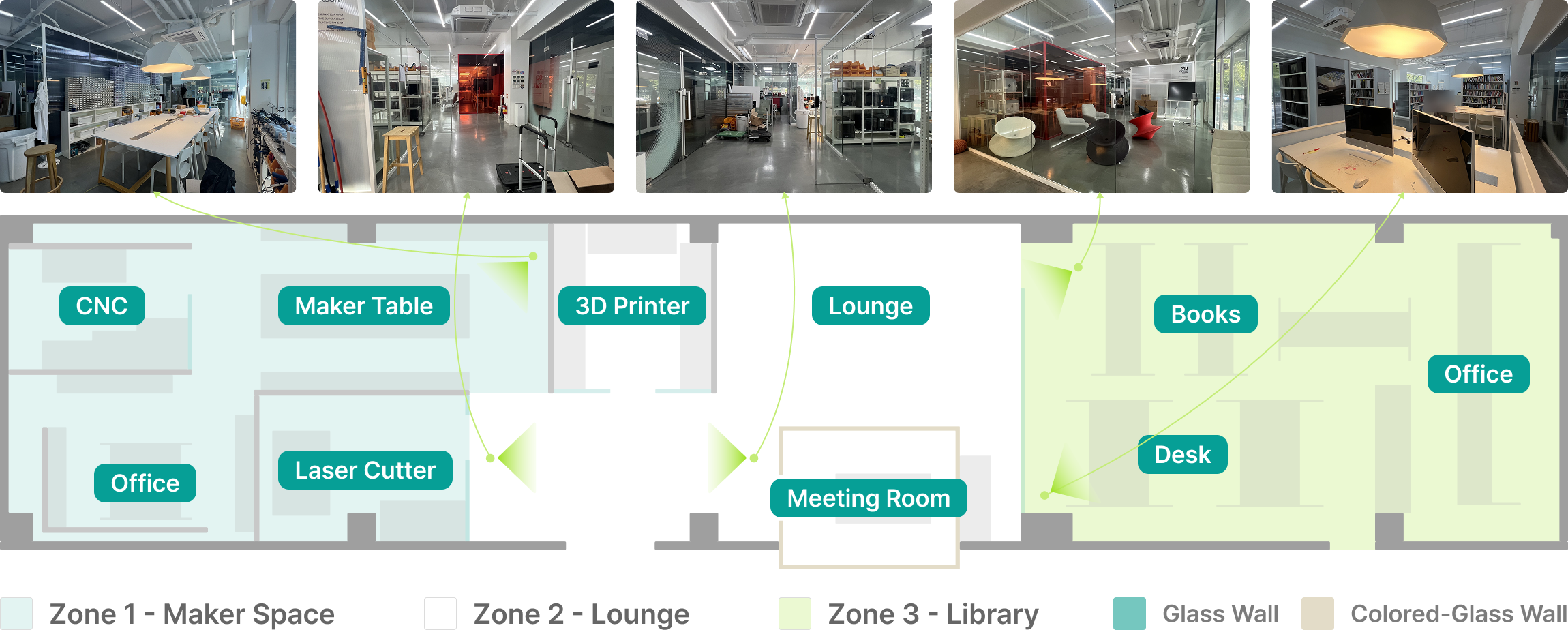}
    \caption{Floor plan and zone divisions of the formative study environment.}
    \Description{A floor plan divided from left to right into Zone 1, the maker space; Zone 2, the lounge; and Zone 3, the library. Labeled areas include the CNC station, maker table, 3D printer, laser cutter, lounge, meeting room, bookshelves, desks, and offices. Five photographs above the plan show representative views of the environment, with arrows linking each photograph to its location. Glass and colored-glass walls are marked separately.}
    \label{fig:environment}
\end{figure}

\begin{figure}[h!]
    \centering
    \includegraphics[width=0.9\linewidth]{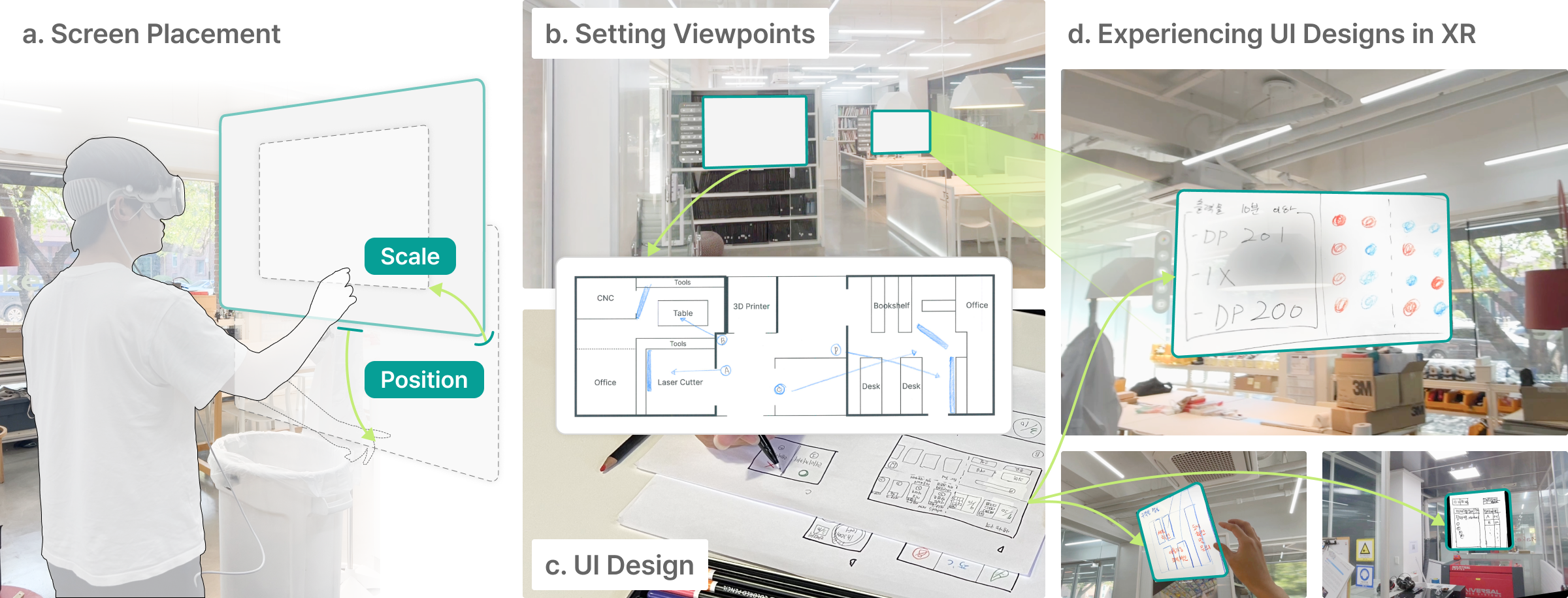}
    \caption{Formative study task process: \textbf{a)} Participants use HMD and browser to choose the position, scale, and angles of the screens, \textbf{b)} Set the viewpoints for each screen and document them on the worksheet, \textbf{c)} Design the UI on paper using pens and pencils, \textbf{d)} Review the UI designs in the XR environment and iterate.}
    \Description{A four-part photographic sequence. Panel a shows a headset-wearing participant positioning and scaling a virtual screen. Panel b shows virtual screens in the study environment and their viewpoints marked on a floor-plan worksheet. Panel c shows paper interface sketches being created. Panel d shows the sketches displayed as virtual panels at several physical locations. Green arrows connect the stages from placement and viewpoint selection to paper design and in-context review.}
    \label{fig:design-task}
\end{figure}

We ran the study in a maker's lounge measuring 33m $\times$ 7.2m with three functionally distinct zones---fabrication equipment, a communal lounge, and a library area (Figure~\ref{fig:environment})---following established behavioral-mapping protocols for spatial design \cite{Sommer_2001} informed by proxemic interaction research \cite{Ballendat_2010}.
The zone separation set three viewing distances: within-zone (average 2.8m), adjacent-zone (approximately 11m), and non-adjacent-zone (approximately 22m).
Each two-hour session opened with consent, a short demographic interview, and a five-minute briefing on the study aim, followed by a headset tutorial on placing, scaling, and rotating a window.
Participants then spent ten minutes placing at least two virtual screens in each of two zones, and every window had to remain visible from at least one point in each zone.
Three design rounds of about twenty minutes followed, one from the central lounge and one from inside each of the two other zones, so every screen was designed from within its zone, from the adjacent zone, and from the non-adjacent zone.
In each round, participants sketched interfaces on paper with real text content, we photographed and displayed the sketches as virtual screens on Vision Pro, and participants reviewed them in place, adjusted window size or angle where needed, recorded their viewpoint and screen positions on a floor-plan worksheet, and iterated before the next round (Figure~\ref{fig:design-task}).
After the third round, participants placed their designs on the screens and stated the distances at which they wanted each design to switch.

We recruited twelve designers (7 female, 5 male; $M = 2.5$ years of UI design experience, $SD = 1.4$) with design-related academic backgrounds, HMD familiarity, and prior XR use (Appendix~\ref{appendix:demographic}).
Only two had authored XR interfaces before, so the sample reflects the screen-interface practice \systemName targets.
Each session closed with a 25-minute semi-structured interview on design choices, spatial influences, and adaptation challenges (Appendix~\ref{appendix:interview}); participants received approximately 40 USD in local currency.
We collected audio, interface designs, headset camera views, and floor-plan markings of viewpoints and screen positions, and we analyzed 148 interface designs across 48 screen placements together with the transcripts through the six-phase thematic analysis \sout{of Braun and Clarke }\cite{Braun_2006}.
Three authors coded transcripts in English alongside the designs and floor plans, first in one joint pass and then in rotation, so that each author layered codes over every session; the three code sets were merged by discussion into the themes reported below.
Each design was coded as one unit and tagged with the adaptation strategies visible in it, and these tags yield the proportions (Figure~\ref{fig:quant}).

\subsection{Finding 1: Adapting UIs on Element Level}

\begin{figure}[h!]
    \centering
    \includegraphics[width=0.9\linewidth]{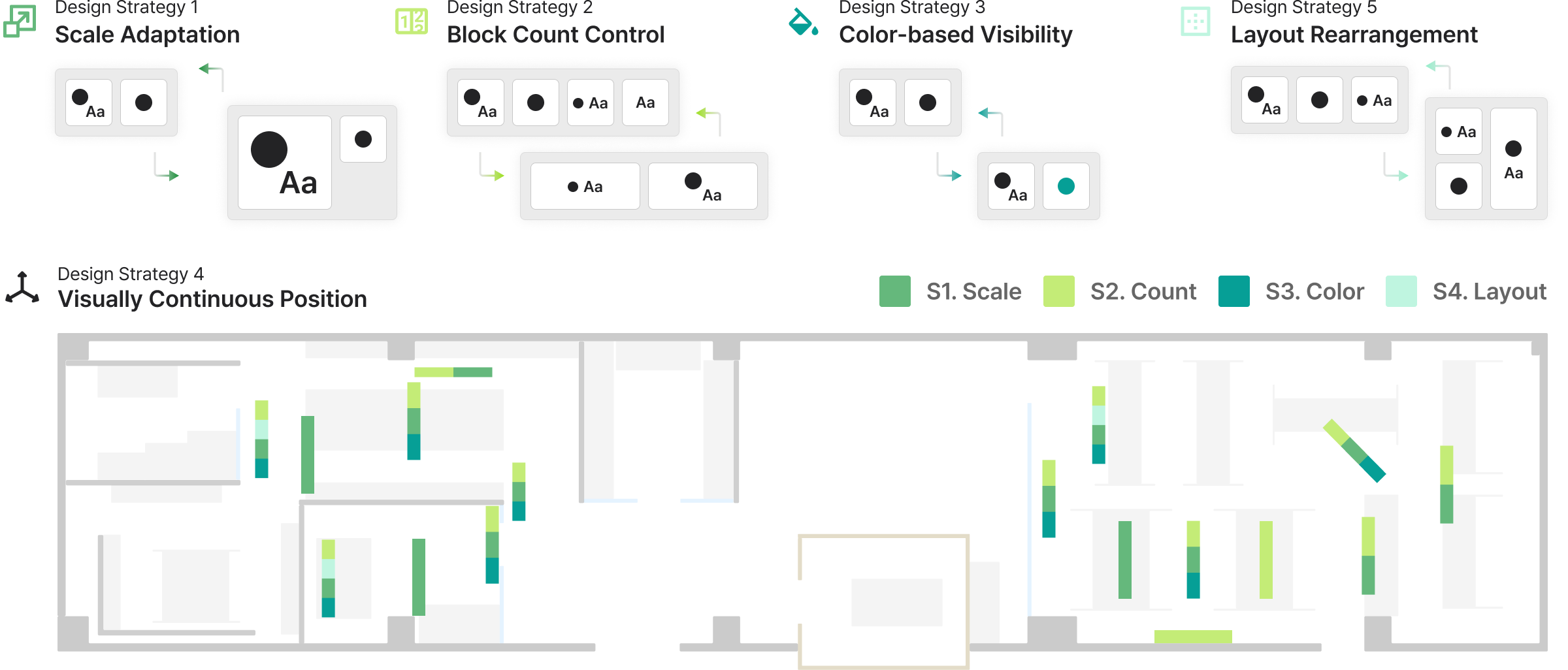}
    \caption{Five key strategies for UI design, along with a floor plan indicating where each strategy is applied.}
    \Description{The upper portion uses before-and-after interface diagrams to illustrate scale adaptation, block-count control, color-based visibility, and layout rearrangement. The lower portion presents visually continuous positioning and maps observed uses of all five strategies onto interface locations throughout the study environment. Colored bars identify the strategies observed at each location.}
    \label{fig:strategies}
\end{figure}

\begin{figure}[h!]
    \centering
    \includegraphics[width=0.9\linewidth]{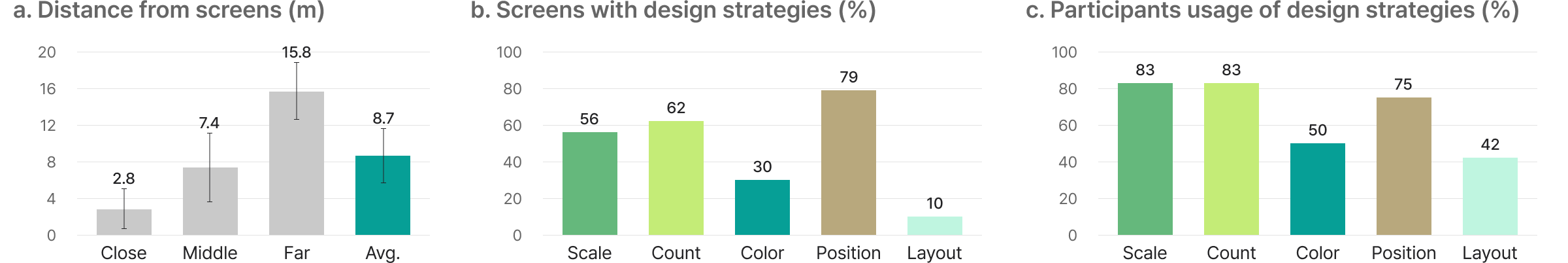}
    \caption{UI design strategy statistics: \textbf{a)} Distribution of distance by zone division, \textbf{b)} Distribution of design strategies by screen, \textbf{c)} Distribution of design strategies by participant.}
    \Description{Three bar charts. Panel a reports mean viewing distances of 2.8 meters for the close zone, 7.4 meters for the middle zone, and 15.8 meters for the far zone, with an overall mean of 8.7 meters. Panel b shows the percentage of screens using each strategy: scale 56 percent, block-count control 62 percent, color 30 percent, position 79 percent, and layout 10 percent. Panel c shows the percentage of participants using each strategy: scale 83 percent, block-count control 83 percent, color 50 percent, position 75 percent, and layout 42 percent.}
    \label{fig:quant}
\end{figure}

Adaptation happened at the level of individual elements, through five recurring strategies (Figure~\ref{fig:strategies}; distributions in Figure~\ref{fig:quant}).
\textbf{Scale adaptation} appeared on 56\% of screens (83\% of creators): participants enlarged text as distance grew, and 42\% of creators swapped text for shapes and icons once even large text stopped reading---P12: \textit{"Once you are far enough away that even large text does not convey meaning, it is better to use shapes to communicate the information."}
\textbf{Block count control} appeared on 62\% of screens (83\% of creators): P6 reduced one design from four blocks at 2.5m to two blocks at 14.7m, and P4 explained, \textit{"When far away, the maximum visible blocks should be limited to two or three. At a distance, I kept only one key component set centered."}
\textbf{Color-based visibility} served binary status at range (30\% of screens, 50\% of creators)---P3: \textit{"I intended green glowing to indicate presence, making it more noticeable."}
\textbf{Visually continuous positioning} governed placement: 75\% of creators avoided occluding physical objects, and 79\% of screens sat against walls or shelves for a flat backdrop.
\textbf{Layout rearrangement} stayed rare (10\% of screens), because participants feared it would read as a different screen---P8: \textit{"If component positions change dramatically or components appear and disappear, users perceive it as a completely different screen, requiring additional mental energy to relearn."}

These strategies form a compact vocabulary of element-level outputs---resize, reduce, recolor, reposition, reorder---applied per condition, yet every adapted state existed only as another static copy whose triggering condition lived in the designer's memory, as in the duplicated artboards of current practice \cite{KraussVRST2021, XRProfessionalInterview}.
This finding motivates \textbf{DG1: element-targeted rule authoring}---let designers select the concrete elements they already think in and state the condition--output mapping as an explicit, editable rule.

\subsection{Finding 2: Deciding Which Versions to Focus}

Designers found it hard to decide which conditions deserved a distinct version and to treat the same condition consistently across versions.
Participants authored progressive series---P12 re-sized text at 15\%, 26\%, then 41\% of the window across three distances---and prioritization strained as versions multiplied; P8: \textit{"It becomes difficult to decide which components to focus on first when there are too many UI windows. Users must judge what to focus on, which takes mental energy."}
The same condition also drew inconsistent treatments: P12 color-coded seat availability at 5.0m and 11.0m but dropped color at 15.8m, while P6 set the threshold elsewhere: \textit{"When I'm far away, I just want to know which ones are in use. I don't think it necessarily needs to be color-coded."}
Participants were further unsure which situations deserved a version at all; P10: \textit{"Should I know information about that space when I'm not entering it?"}

The task fixed how many versions each screen received, but which conditions each version served, and whether sibling versions agreed, was the designers' own work, and nothing in the designers' materials showed which conditions were handled and which were missed.
This finding motivates \textbf{DG2: multi-condition preview comparison}---render each rule's condition-specific state beside the default so coverage gaps and mismatched siblings become visible at a glance.

\subsection{Finding 3: Judging Transitions in Place}

The decisions participants found hardest---when an adaptation should fire and how it should change---were the ones they could not settle away from the space.
Several preferred discrete, zone-based switching; P12: \textit{"From the user's perspective, thinking there's something physical there, and when they get closer it gets bigger---this feels more naturally acceptable."}
Others feared jarring changes near boundaries---P7: \textit{"When I get close, the screen suddenly changes, that would be very confusing"}---and P11 resisted continuous change: \textit{"If the size changes every time I move, it feels chaotic."}
Occlusion problems, likewise, surfaced only through physically moving in the space; P11: \textit{"But when this wall blocks like this, this icon gets hidden, so I was wondering what to do about that,"} and P8 was \textit{"forced to place it here"} by a bookshelf that blocked or blurred the view.

Participants could author adapted states on paper, but validating a trigger point or a transition required standing where the interface would live.
This finding motivates \textbf{DG3: situated testing of the active rule}---let designers dial or sense the environmental conditions in place and watch the one currently matched rule drive the design.
Of \systemName's three sensed conditions, distance and occlusion emerged directly from this study, and lighting was added as a condition from prior findings on legibility in dim and bright environments \cite{DarkOSTHMD_2019, DarkOSTHMD_2021}.

\section{\systemName}

\begin{figure}[h!]
    \centering
    \includegraphics[width=1\linewidth]{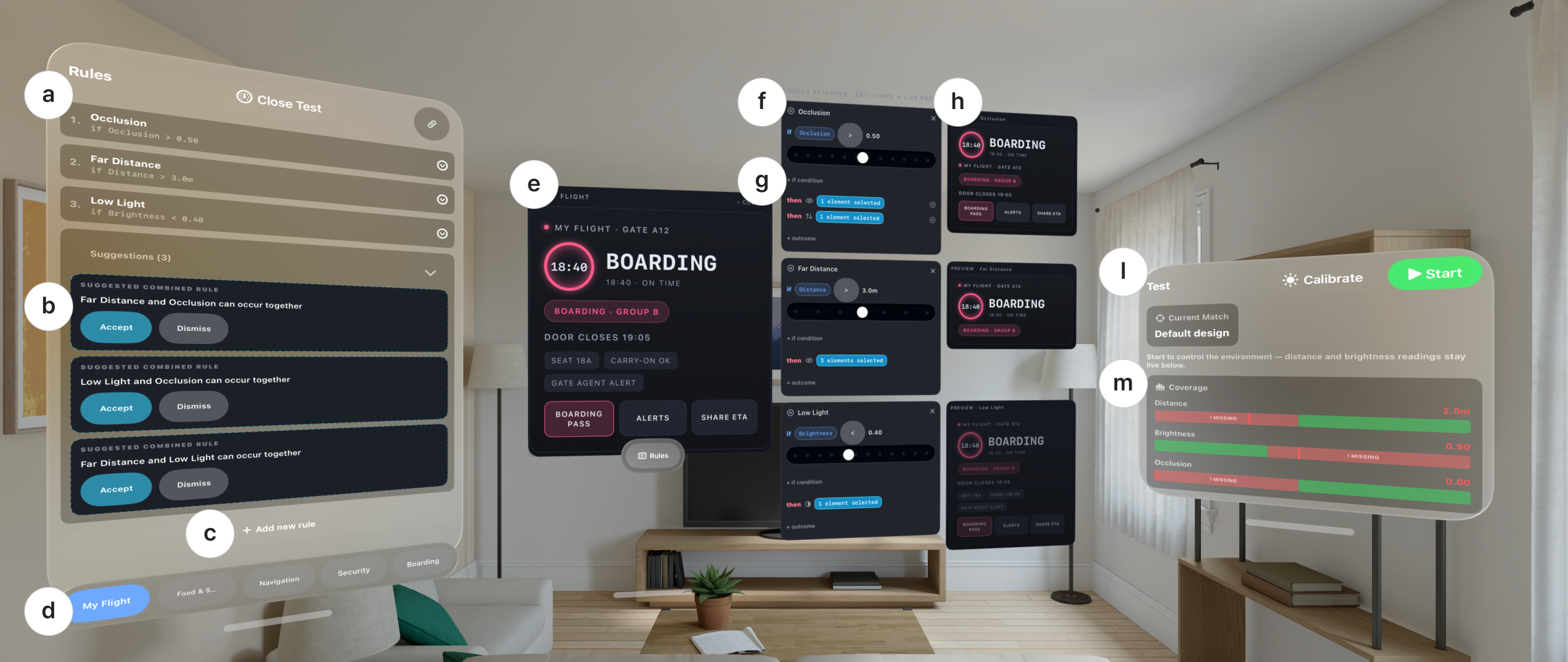}
    \caption{\systemName overview: Users can \textbf{a)} View all created rules and reorder to determine priority. \textbf{b)} Get suggestions for combining rules for different scenarios. \textbf{c)} Add new rules. \textbf{d)} Show or hide for each imported UI. \textbf{e)} Review the original UI and select elements for rule application. \textbf{f)} Set environmental triggers (IF statements) with specific values. \textbf{g)} Style output for the THEN actions. \textbf{h)} Preview output of created rules. \textbf{l)} Test to see how adaptive UIs respond to manual or \textbf{m)} Live sensor values like distance, brightness, and occlusion.}
    \Description{A spatial view of \systemName arranged as floating panels in a living room. The Rules panel on the left lists prioritized rules and suggested combinations. The center contains an imported flight-status interface beside three rule editors, each paired with an adaptive preview. The Test panel on the right shows the current matching design, a Start control, and coverage strips for distance, brightness, and occlusion. Lettered callouts identify the corresponding controls and panels.}
    \label{fig:system_overview}
\end{figure}

\systemName is an authoring tool for designing adaptive interfaces in the environments where they will be deployed (Figure~\ref{fig:system_overview}). It lets designers author adaptive behavior as rules over interface elements, compare the condition-specific versions side by side, and test which becomes active directly in the intended physical environment. The resulting design is a set of adaptive interface variants together with the rules that define when each adaptation should occur.

The tool was structured in response to the findings of the literature review and formative study, and designed with three goals: DG1) let designers author adaptation rules directly on the elements of their imported design, stating \textbf{what should change and under which environmental condition}; DG2) let designers \textbf{compare each condition-specific variant} beside the default design, so overlapping and uncovered conditions stay visible; and DG3) let designers \textbf{test which variant becomes active in the real environment}, under sensed or manually controlled values, and refine the rules in place.
A typical session runs as follows. A designer imports a multi-screen design, places its screens around the room, and taps an element to make it a rule's target (DG1). The designer sets a condition, such as \emph{distance} \textgreater{} \emph{3\,m}, adds an outcome, such as a larger size, and watches the preview next to the rule update as they edit (DG1). Expanding the screen shows a preview of every rule next to the default; a suggestion card notes that two of their rules can fire together, and the coverage strip shows that brightness has no rule yet, so they add one (DG2). They then start Test Mode, walk across the room, and watch the far variant appear, drag the distance slider past the room's far wall to check a condition the space cannot reach, stop, nudge the threshold, and test again (DG3). Once the rules hold, they copy a rule onto the other screens and export the project for handoff.

\subsection{DG1: Element-Targeted Rule Authoring}

\begin{figure}[h!]
    \centering
    \includegraphics[width=1\linewidth]{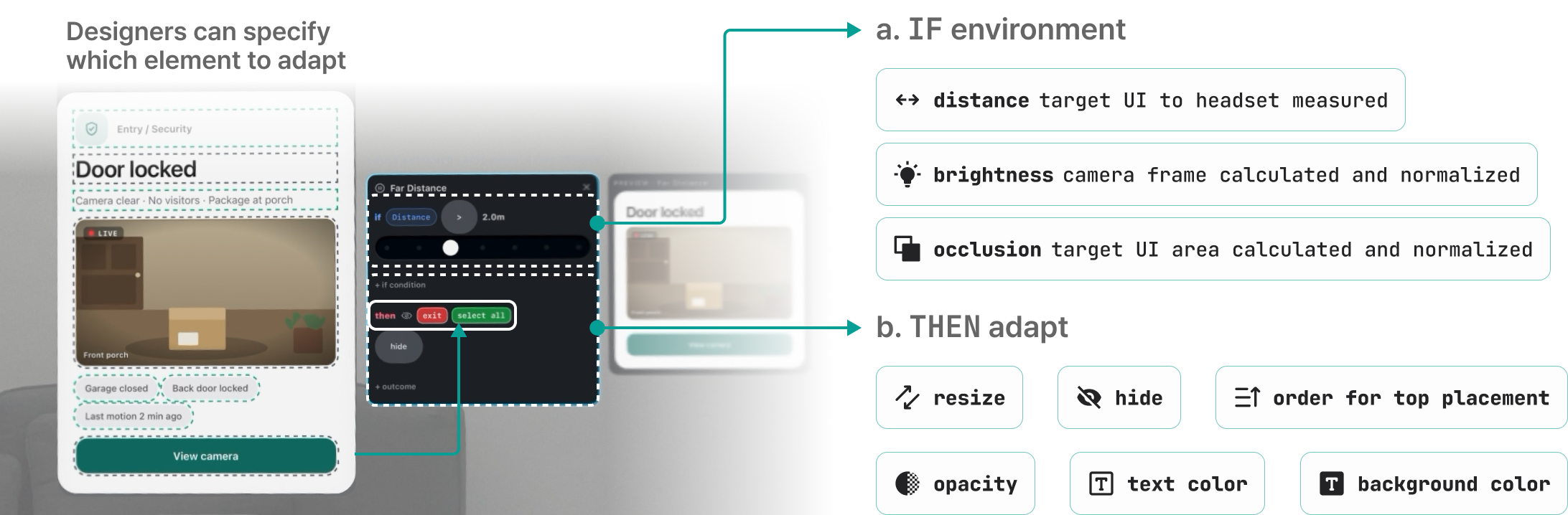}
    \caption{\systemName supports \textbf{a)} IF condition authoring with environmental measures, and \textbf{b)} THEN output authoring effective on the whole screen or selected element(s).}
    \Description{An element-targeted rule appears between an imported security interface on the left and two option groups on the right. Dashed outlines divide the interface into selectable elements, and an arrow links the rule's selected-element control to those targets. The IF group lists distance, normalized brightness, and normalized occlusion. The THEN group lists resize, hide, placement order, opacity, text color, and background color.}
    \label{fig:system_rule}
\end{figure}

Designers author each adaptation directly on the rendered design. Importing an interface opens a spatial window per screen (Figure~\ref{fig:system_overview}e). The designer places these windows around the room, scales one with a two-hand pinch, and toggles each open or closed from the workspace bar (Figure~\ref{fig:system_overview}d); reopening a saved project restores its rules, variants, and test records. To set \emph{what} changes, the designer taps an element (e.g., the BOARDING GROUP B chip). To set \emph{when}, the designer picks a measure, a comparison, and drags a slider to a threshold such as \emph{distance} \textgreater{} \emph{3\,m} (Figure~\ref{fig:system_rule}a). Pressing \emph{+\,outcome} then gives the selected elements a visual change: hiding them or bringing them to the front by selection, stepping size or opacity with a slider, or setting a text or background color with a picker (Figure~\ref{fig:system_rule}b). Each edit plays back in the preview attached to the rule (Figure~\ref{fig:system_overview}h).

Each rule is an IF--THEN pair over the chosen elements, and \emph{+\,if condition} adds conditions so one rule can describe a compound situation, such as a far viewer in a dim room or a 2 to 4 meter distance band. Distance and occlusion follow our formative study; brightness follows legibility findings on dark-mode interfaces in optical see-through head-mounted displays~\cite{DarkOSTHMD_2019, DarkOSTHMD_2021}. A THEN output applies to the selected elements or to the whole screen, and one rule may carry several outputs, so different elements respond differently under the same condition, and the output palette (hide, bring-to-front, resize, opacity, text and background color) follows the adaptation strategies designers used in the formative study. When a rule activates, its elements ease into the new state over a short transition, so a hidden element fades and a resized one grows into place. The Rules panel lists the rule set for renaming, enabling, disabling, duplicating, and deleting (Figure~\ref{fig:system_overview}a); when several rules match at once, the highest in the list takes effect, and designers reorder rules with the arrows on each card. On the web editor, designers can also drag preset condition cards onto a screen or element to seed a rule.

Designers can carry a finished rule onto another screen. Pressing \emph{apply to} lists the project's other screens and copies the rule onto the chosen one (Figure~\ref{fig:system_overview}a); where the original target is missing on that screen, the designer picks which elements the copy should affect.

\subsection{DG2: Multi-Condition Preview Comparison}

As rules accumulate, designers need to see how the variants relate: where conditions overlap, clash, or leave a situation unaddressed. Expanding a design window lays each rule's preview beside the unchanged default, so several variants read side by side (Figure~\ref{fig:system_overview}h). When two rules could fire together, a suggestion card appears(\emph{``Far Distance and Occlusion can occur together''}), and the designer accepts it to merge the rules or dismisses it (Figure~\ref{fig:system_overview}b). When two rules set the same property to different values, a conflict card appears, and the designer combines the rules, chooses which wins, or dismisses it.

The coverage strips which ranges rules already cover and which remain open, with a marker at the live value (Figure~\ref{fig:system_overview}l, m). Tapping an open range closes it, and the designer chooses whether the design eases gradually into the covered state or snaps to it; a measure with no rule at all prompts a suggestion to add one. In the web editor, a designer can also select two variants and view a per-layer list of all properties that differ between them. Every preview updates as its rules' conditions, targets, or outputs change, so overlaps, conflicts, and gaps stay visible on one authoring surface.

\subsection{DG3: Situated Testing and Refinement}

\begin{figure}[h!]
    \centering
    \includegraphics[width=0.45\linewidth]{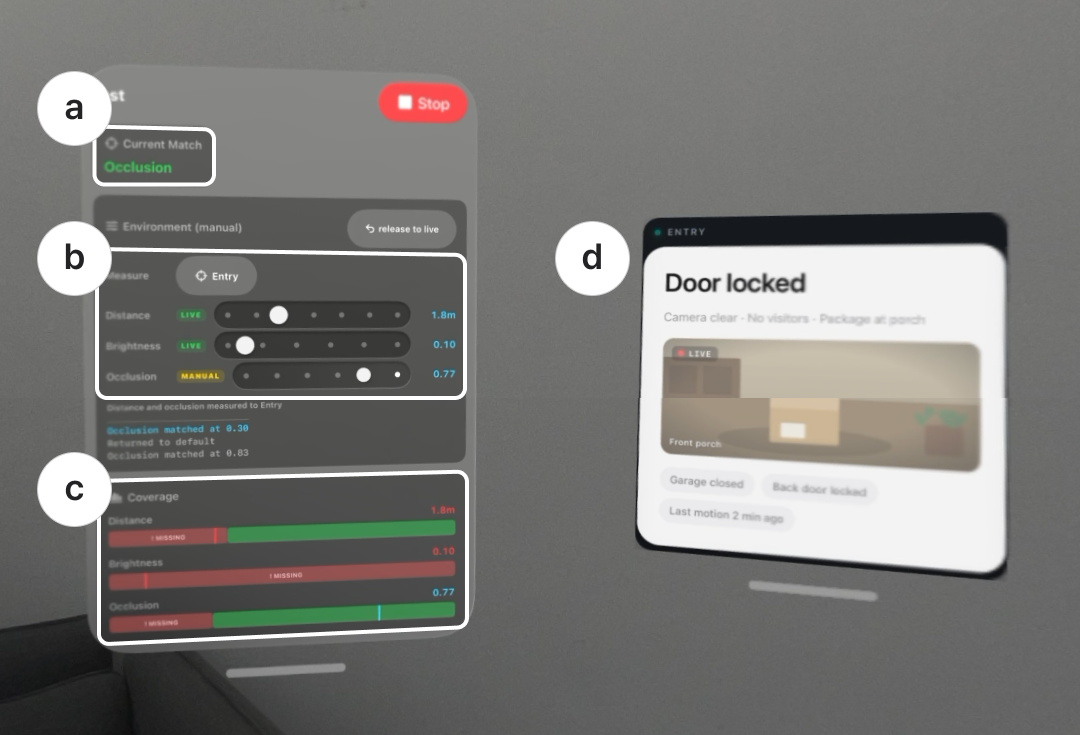}
    \caption{\systemName Test Mode: \textbf{a)} currently activated design version, \textbf{b)} current environmental values from live sensing or manual control, \textbf{c)} coverage strip summarizing addressed and under-addressed environmental factors, and \textbf{d)} interfaces responding to active sensor values.}
    \Description{A headset view containing a Test panel and an entry-security interface. The Test panel identifies occlusion as the current match. Distance and brightness are live, while occlusion is manually controlled; their displayed values are 1.8 meters, 0.10, and 0.77. Coverage strips use green to indicate addressed ranges and red to indicate missing ranges. The entry-security interface on the right displays the resulting adapted state.}
    \label{fig:system_test}
\end{figure}

Judging whether an adaptation is right, such as what counts as ``far'' or whether text stays legible in the available light, requires being in the space. Test Mode therefore evaluates authored rules in the physical setting where the interface will appear.
Pressing \emph{Start} on the Test panel folds the editing controls away, and the screens begin reacting to the room: stepping closer, moving into shadow, or blocking a screen with a hand switches it to the matched variant (Figure~\ref{fig:system_test}). The panel names the active rule, shows the live distance, brightness, and occlusion values, and keeps a trace of recent switches (Figure~\ref{fig:system_test}a, b). A \emph{Calibrate} control re-baselines brightness to the current room, and a \emph{Measure} menu sets which screen the readings target.

When the room cannot reach a \sout{value---a}\jw{value, such as a} viewing distance longer than the \sout{space allows---the designer can drag that sensor's slider on the Test panel}\jw{space, the designer drags that measure's slider} to set it by \sout{hand, while the other measures}\jw{hand while the others} keep reading from the \sout{room (Figure~\ref{fig:system_test}b). Letting go of}\jw{room; releasing} the slider returns \sout{that measure to live sensing. This lets the designer manually preview a rule at conditions that are hard to stage in the available space while keeping the rest of the sensed context intact.}\jw{it to live sensing (Figure~\ref{fig:system_test}b).}
\sout{\systemName runs Test Mode from live readings evaluated on the headset. It senses distance, brightness, and occlusion for every open screen each frame and evaluates that screen's enabled rules against the readings, so}\jw{Screens follow} the highest-priority matching rule \sout{selects the shown variant and the default returns when none matches. A newly matched variant takes effect once it has stayed the best match through a brief settling window, which keeps a reading that hovers}\jw{and return to the default when none matches, and a new match takes effect only after it holds briefly, so a reading hovering} near a threshold \sout{from flickering between states.}\jw{does not flicker.}
Leaving Test Mode returns the designer to authoring, where changing a threshold, target, output, or priority, and then pressing \emph{Start} again, repeats the test.

Once \sout{designers are satisfied, they can export}\jw{satisfied, the designer exports} the project for continued design work, developer handoff, or sharing. \sout{Because \systemName is a design-time tool, the}\jw{The} export is HTML/CSS \sout{along} with the rule and style specifications, so a developer can implement the behavior \sout{in the running product} after designers have iterated \sout{on it in context. The}\jw{it in context, and the} saved project keeps the default interface, \sout{the authored variants, the}\jw{variants,} specifications, and test \sout{records, so the adaptive logic travels with the design beyond \systemName.}\jw{records.}

\subsection{System Implementation}

\begin{figure}[h!]
    \centering
    \includegraphics[width=1\linewidth]{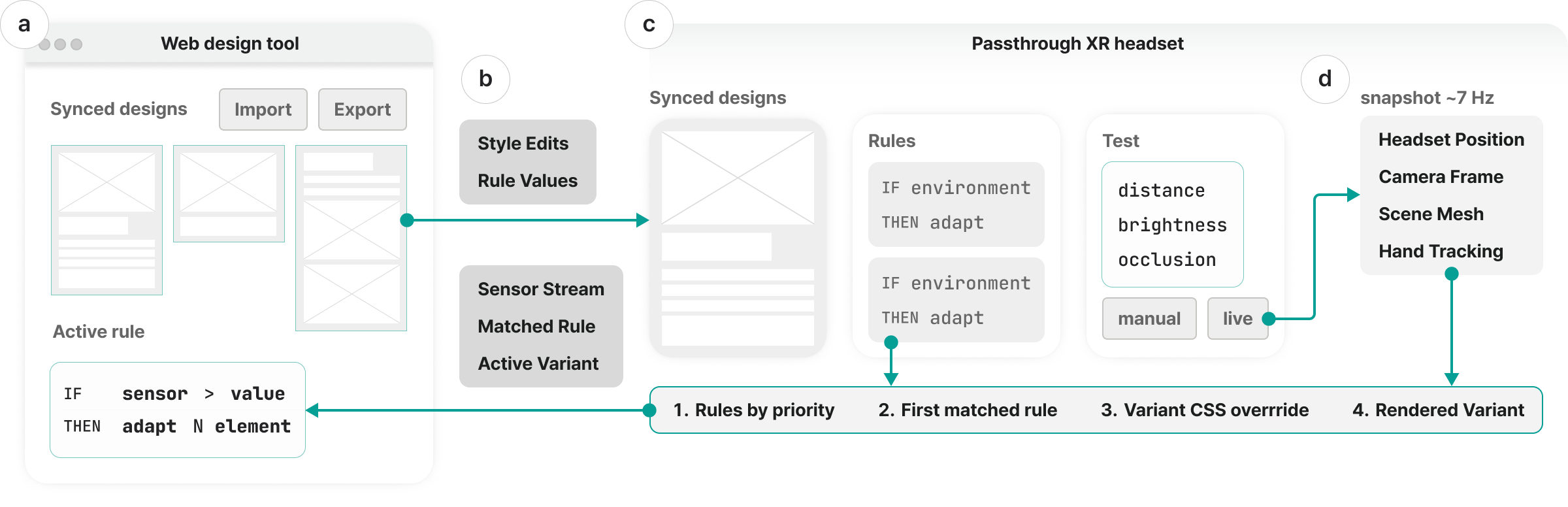}
    \caption{\systemName architecture: \textbf{a)} The React web editor \textbf{b)} exchanges a shared project through a typed message bridge with \textbf{c)} a SwiftUI visionOS application. Both surfaces render the same HTML/CSS source, while \textbf{d)} the active rule engine applies variant-specific CSS during testing.}
    \Description{A left-to-right architecture diagram. The web design tool imports, edits, and exports synchronized designs. A bidirectional bridge sends style edits and rule values to the headset and returns sensor data, the matched rule, and the active variant. The headset contains synchronized designs, IF--THEN rules, and manual or live controls for distance, brightness, and occlusion. Live values are derived at approximately 7 hertz from headset position, camera frames, the scene mesh, and hand tracking. The rule engine evaluates rules by priority, selects the first match, applies its CSS override, and renders the resulting variant.}
    \label{fig:system_architecture}
\end{figure}

\systemName combines a React and TypeScript web client with a native SwiftUI application for visionOS (Figure~\ref{fig:system_architecture}). The web client handles \sout{interface} import, project management, and detailed rule \sout{editing, while} \jw{editing;} the native application presents the rules panel, Test panel, and imported interfaces as \sout{independent} spatial windows. Both \sout{applications operate within a shared project model that includes interface} \jw{share one project model of} screens and layers, \sout{adaptive variants, environmental} \jw{variants,} rules, window placements, and test sessions.

\sout{When an HTML/CSS interface is imported,}\jw{On import,} the web client extracts \sout{its}\jw{an interface's} layers as CSS selector paths and creates a default variant \sout{for each screen. Each}\jw{per screen; each} adaptive variant stores only the CSS properties that differ from \sout{this default, while its associated rule records the environmental}\jw{the default, and its rule records} conditions, target variant, enabled state, and priority. Both applications render the same HTML/CSS \sout{source through \texttt{WKWebView}, with variant-specific}\jw{through \texttt{WKWebView} with the variant's} CSS injected over \sout{the default interface to keep authoring previews and spatial windows visually consistent.}\jw{it.} A typed JavaScript-to-Swift bridge synchronizes project updates, rule edits, \sout{window} placements, sensor readings, and active-rule changes as JSON \sout{messages. Each stateful update includes its}\jw{messages; updates carry a} source and timestamp \sout{for last-write-wins conflict resolution, and native edits are queued during temporary disconnections before being replayed when the connection is restored.}\jw{so the latest write wins, and edits made while disconnected replay on reconnection.}

Rule evaluation \sout{is implemented}\jw{runs} in both TypeScript \sout{and Swift}\jw{(authoring previews) and Swift (Test Mode)} with the same priority behavior. 
\sout{Each frame, the enabled rules of every open screen are evaluated}\jw{On the headset, a sensor loop runs every 150\,ms, evaluates each open screen's enabled rules} against its current \sout{viewer} distance, brightness, and \sout{occlusion. The matching rule with the lowest-priority number selects the displayed variant, and the default interface is restored when no rule matches. To keep a reading that hovers at a threshold from flickering, a}\jw{occlusion, and shows the variant of the lowest-numbered matching rule, restoring the default when none matches. A} newly matched variant commits only after it stays the best match \sout{through a short settling window. Variant changes are applied by patching}\jw{for 300\,ms, and variant changes patch} the variant's CSS into the live document with a \sout{brief eased transition, including discrete properties such as visibility, so both authoring previews and Test Mode animate between states. The TypeScript evaluator updates authoring previews, while the Swift evaluator controls Test Mode directly on the headset. Combined-rule suggestions are generated by identifying compatible condition overlaps and merging their CSS outputs. Conflicting}\jw{400\,ms eased transition. Suggestions and coverage strips derive from the same rules: each rule's conditions reduce to a value range per measure; compatible overlaps become combined suggestions that merge both rules' CSS, conflicting} properties are surfaced for \sout{designer resolution, and cross-screen}\jw{the designer, and the strips mark the covered and open bands of each measure. Cross-screen} reuse maps source elements \sout{through semantic roles before designers select the corresponding target elements and create a screen-specific copy.}\jw{by semantic role (primary content, secondary content, status, action button) before the designer confirms targets and a screen-specific copy is created.}

Live \sout{environmental values are supplied through \texttt{ARKit}. World tracking measures the distance between the headset and each interface window, adjusted by a fixed calibration offset,}\jw{values come from \texttt{ARKit}. Distance is the world-tracked distance from the headset to each window,} with a geometry-based \sout{estimate as a} fallback when tracking \sout{is unavailable. Camera luminance provides the brightness estimate, with environment light estimation as a fallback and a designer-triggered recalibration that re-baselines it to the room. Occlusion is estimated by sampling a grid of points across each window's world geometry. Tracked hands take precedence,}\jw{drops. Brightness comes from environment light estimation, normalized between the lowest and highest values seen since launch or the last \emph{Calibrate}; on devices with main-camera access, the camera frame's mean luminance is used directly. Occlusion samples a 3$\times$3 grid on each window and scores each point by how close a tracked hand comes to the line from the headset to that point (fully occluded within 0.10\,m, clear beyond 0.30\,m),} and ray tests against the \sout{reconstructed scene mesh are filtered by ARKit surface classification, so}\jw{classified scene mesh add} furniture such as \sout{a table or seat counts as an occluder. Live measurements and manual}\jw{tables and seats as occluders. Manual} overrides enter the same \sout{rule evaluator. For export, \systemName combines the base interface with each variant's CSS overrides to produce static HTML artifacts. A complete JSON package preserves the project model}\jw{evaluator. Export writes each variant as static HTML with its CSS overrides, a JSON package of the project} and test sessions, \sout{while separate}\jw{and} rule and style specifications \sout{support inspection and developer handoff. Exported}\jw{for handoff;} variant HTML can \sout{also be returned}\jw{return} to Figma through an external HTML-to-design \sout{plugin; \systemName does not provide native Figma synchronization.}\jw{plugin.}

\begin{table}[H]
\footnotesize
\centering
\caption{Observable end-to-end adaptation latency from video annotation. Each condition uses the first 100 valid environmental-change/UI-response pairs.}
\label{tab:latency}
\scalebox{0.8}{%
\begin{tabular}{lrrrr}
\toprule
Condition & Median (s) & Mean (s) & IQR (s) & Range (s) \\
\midrule
Distance   & 0.77 & 0.81 & 0.67--0.93 & 0.37--1.57 \\
Brightness & 1.70 & 1.81 & 1.07--2.43 & 0.23--4.50 \\
Occlusion  & 0.37 & 0.46 & 0.30--0.57 & 0.10--0.97 \\
\bottomrule
\end{tabular}}
\end{table}

\sout{To characterize the responsiveness of live testing, we measured observable}\jw{We measured} end-to-end adaptation latency for \sout{distance, brightness, and occlusion. For each environmental factor, we recorded headset-view videos while repeatedly changing the relevant condition and}\jw{each factor from headset-view video. We repeatedly changed one condition,} annotated the first 100 valid pairs of \sout{events: the}\jw{a} visible environmental change and the first visible UI \sout{adaptation response. Latency was computed as the time difference between these annotations, capturing}\jw{response, and used the time between them as latency. This covers} sensing, rule evaluation, synchronization, \sout{rendering, and visible feedback as experienced by the designer (Table~\ref{tab:latency}); brightness}\jw{and rendering as the designer experiences them (Table~\ref{tab:latency}). Brightness} was slower and more variable because \sout{physical lighting transitions and camera luminance stabilization introduced additional delay.}\jw{room lighting changes more slowly and the light estimate settles gradually.}
\section{User Study}

We ran a within-subjects study comparing how designers author adaptive XR interfaces with \systemName against their current 2D practice.
\systemName offers three capabilities---element-targeted rules, rule-attached previews with combination suggestions, and in-context Test Mode---and we ask how \sout{each one changes a concrete part of} designer behavior \jw{on three fronts differs between \systemName and} \sout{relative to} a Figma-based \jw{current-practice} baseline:
\textbf{RQ1.} \sout{Do element-targeted rules change how}\jw{How} explicitly \sout{designers}\jw{do designers} specify adaptive behavior---which element changes and under which condition---\sout{rather than leaving it}\jw{and where does that specification live, in a working rule or} in text or speech?
\textbf{RQ2.} \sout{Do rule-attached previews and combination suggestions change how}\jw{How} many environmental conditions, including compound ones, \sout{a designer covers}\jw{do designers cover}?
\textbf{RQ3.} \sout{Does in-context Test Mode change where thresholds and styling come from---values guessed at a desk versus values observed in the headset}\jw{How do designers test their adaptations in place, and how do tests feed back into thresholds and styling}?

\subsection{Participants}
We recruited twelve designers (7 female, 5 male; ages 20--31, $M = 25.0$, $SD = 2.8$) via university community channels.
Participants included UI/UX, industrial design, and HCI researchers and graduate students in related domains. They had an average of 3.2 years of design experience ($SD = 1.7$; range: less than one year to more than five years).
Eligibility criteria included prior UI design practice, at least one experience with an HMD or AR/VR, normal or corrected-to-normal vision, and intact color vision. Seven participants also reported XR UI design experience through coursework or research projects (Table~\ref{tab:study_participants}).
All twelve used Figma as a primary design tool and reported high Figma fluency ($M = 6.0$ of 7, $SD = 1.0$); thus, the baseline condition used a tool with which they were already proficient.
Participants received approximately 35 USD in local currency for the two-hour session; the protocol was approved by our institutional review board.

\begin{table}[h!]
\centering
\caption{Self-reported user study participant demographics}
\label{tab:study_participants}
\setlength{\tabcolsep}{4pt}
\resizebox{\linewidth}{!}{%
\begin{tabular}{l c c l c l c c l}
\toprule
\textbf{ID} & \textbf{Gender} & \textbf{Age} & \textbf{Background} &
\makecell{\textbf{Design}\\\textbf{Exp.}} &
\textbf{UI Design Tools} &
\makecell{\textbf{Figma-like Tool}\\\textbf{Fluency (1--7)}} &
\makecell{\textbf{XR Familiarity}\\\textbf{(1--7)}} &
\textbf{XR UI Design Exp.} \\
\midrule
P1  & M & 24 & Computer Science      & $<$1 yr  & Figma, code        & 5 & 3 & -- \\
P2  & F & 25 & Industrial Design     & 4--5 yrs & Figma              & 7 & 2 & -- \\
P3  & F & 24 & Industrial Design     & 4--5 yrs & Figma, code        & 6 & 4 & AR application \\
P4  & M & 20 & Industrial Design     & 1--2 yrs & Figma              & 7 & 2 & AR goggle UI \\
P5  & M & 31 & Computer Science      & 4--5 yrs & Figma, code        & 5 & 4 & -- \\
P6  & F & 26 & Industrial Design     & 3--4 yrs & Figma              & 7 & 5 & -- \\
P7  & F & 26 & Industrial Design     & 5+ yrs   & Figma              & 6 & 5 & VR navigation UI \\
P8  & M & 25 & HCI (XR)              & 1--2 yrs & Figma, code, Unity & 4 & 6 & XR gesture-UI research \\
P9  & F & 25 & Industrial Design     & 1--2 yrs & Figma, Framer      & 6 & 6 & VR simulation UI \\
P10 & F & 21 & Industrial Design     & 2--3 yrs & Figma, code        & 6 & 4 & VR concept UI (Quest) \\
P11 & M & 27 & UX Research/UI Design & 4--5 yrs & Figma, code        & 7 & 5 & AR application (Quest) \\
P12 & F & 26 & Industrial Design     & 4--5 yrs & Figma              & 6 & 5 & -- \\
\bottomrule
\end{tabular}}
\Description{Demographics table for 12 user-study participants. Participants aged 20 to 31, with 7 females and 5 males. Backgrounds include Industrial Design (8 participants), Computer Science (2), HCI, and UX Research. UI/UX design experience ranges from under 1 year to over 5 years, with 6 participants at 4 or more years. All twelve use Figma as a primary prototyping tool; five also code in HTML/CSS/JS, one also uses Unity, and one also uses Framer. Self-rated Figma fluency ranges from 4 to 7 on a 7-point scale (mean 6.0); self-rated XR device familiarity ranges from 2 to 6. Seven participants report some prior XR UI design experience from course or research projects; five report none.}
\end{table}

\subsection{Study Procedure}
Two authors attended each session: one moderated, while the other managed logging, observation, and assistance.
Upon providing consent, participants received an explanation of the research background and completed an intake survey covering demographics and 7-point Likert-scale self-ratings of UI and context-aware design experience, Figma-like tool fluency, and XR usage familiarity.
We then calibrated the Vision Pro's eye and hand tracking for each participant, mirrored the in-headset view to a Mac mini for observation and logging, and provided brief training on moving, resizing, and deleting windows in visionOS.

Each participant then completed the two design tasks in counterbalanced order (six participants per order).
After each task, participants briefly walked us through their artifacts, explaining what they adapted, when each state should appear, and what they changed after previewing or testing. They then completed the post-task survey; administering the debrief and survey after each task reduced recency bias in the end-of-session ratings.
The baseline required no training because participants were already proficient with the tool. 
Before the \systemName task, the authors provided \sout{an approximately 10-minute}\jw{a ten-minute} spoken walkthrough of each capability while observing the mirrored view. This familiarized participants with the system's basic operation before their task behavior was measured.
The session concluded with \sout{an approximately}\jw{a} 15-minute semi-structured interview. Having used both tools, participants compared them, reflected on their own practices, and considered possible uses of \systemName (Appendix~\ref{appendix:study_interview}).
Surveys were administered in English, with verbal clarification in the participant's language upon request.

\subsection{Conditions and Tasks}

The two tasks used matched but distinct domains so that no participant authored the same artifact twice.
\textbf{Task~1 (\systemName)} ran on a smart-home control dashboard (living-room overview, thermostat, entry security) with the full system and suggestions enabled.
\textbf{Task~2 (baseline)} used workplace information panels (team board, meeting-room status, wayfinding) and represented credible current practice~\cite{XRProfessionalInterview}: participants edited on a desktop in Figma, inspected static PNG previews floating on Vision Pro through a minimal view-only image viewer, and returned to Figma between inspections to make revisions.
Each domain comprised three screens built on the same scaffold---a header, a primary reading, a subtitle, one rich element (slider, dial, or image), three chips, and a primary button. The conditions therefore provided structurally equivalent authoring surfaces.
The viewer eliminated the file-transfer friction of everyday practice while holding window chrome, passthrough blending, and logging constant. The baseline was consequently more convenient than the workflow it represented.
\jw{The baseline stands for current Figma-based practice, not for XR authoring tools with conditional-behavior support; it cannot express or execute environmental rules, so rule counts, Test Mode runs, and live triggering are capabilities of one condition only, and we read them as workflow differences}
The tasks had intentionally different allocated time budgets.
Because all participants were fluent Figma users, the familiar baseline received 15 minutes, whereas the unfamiliar \systemName task received 30 minutes. Participants could exceed either budget upon request\sout{ rather than rush to finish}\jw{, so no one had to rush}.
We report behavioral outcomes as time-normalized rates alongside raw totals to account for this difference, and we revisit the condition-to-domain pairing in Limitations.
We address this tradeoff in three ways: by using the matched scaffold described above; by analyzing reasoning-sensitive measures from the first block only, comparing the six \systemName-first participants with the six baseline-first participants; and by reporting the results as \systemName-on-home versus baseline-on-office.

Both tasks used the same brief: \emph{``This panel will be used in [setting]. People will use it in different situations as the day and the space change. Design the three screens so they stay usable and clear for whoever is using them, however the situation changes.''}
The brief did not identify environmental factors because RQ2 and RQ3 examine whether designers notice such conditions independently. Participants worked from provided seed screens, focusing the task on authoring adaptations to an existing design.
We staged the factors without naming them: participants stepped back across the room (distance), the room lights were dimmed (brightness), and participants blocked the design with a hand \sout{or another floating window} (occlusion).
Every participant thus met the same three-staged moments, and whether they addressed each one is a measured outcome.
A task was considered complete when every screen included at least one working adaptive state and the participant had previewed or tested the design at least once.
\jw{The \systemName build used in the study sensed occlusion from tracked hands only, with scene-mesh occlusion disabled so that staging stayed predictable; distance was the head-to-window distance minus a fixed 1.8\,m offset, floored at 0.2\,m, so authored thresholds sit on that working scale; and brightness was ARKit light estimation normalized to the range seen since launch, so the app was launched in the study room before each block.}

\subsection{Data Collection and Analysis}

We logged designers' actions as structured events in both conditions: rules authored and their selected targets (RQ1); coverage of the three staged factors, combined rules, and suggestion accept/edit/dismiss decisions (RQ2); \sout{threshold provenance (sampled, typed, or later edited),}\jw{Test Mode runs, threshold and style edits with their timing relative to tests,} test--refine cycles, and rule reuse across screens (RQ3).
The Figma artifacts of all twelve baseline sessions were retained and coded for variants and their intended conditions. \jw{One author inventoried each file from its pages and layer names together with the walkthrough recording, listing every on-page state as a variant and recording its trigger from a canvas annotation, a layer name, or the participant's spoken explanation; a second author reviewed all twelve inventories, and uncertain cases were settled by discussion. A staged factor counts as covered when at least one variant's trigger names it, and a variant whose trigger names two factors counts as a compound condition. With \systemName, the same counts come from the authored rules, so the baseline criterion is the looser of the two.}

After each task, participants rated raw NASA-TLX (six subscales, 1--10) and SUS; each block additionally carried tool-specific items on 7 point Likert scale---two workflow items in the baseline block (\emph{``I could clearly communicate when the interface should change''}; \emph{``I could directly check whether the adaptation worked in XR''}) and five capability items in the \systemName block (element selection, rules, previews, Test Mode, suggestions).
Qualitative data comprise the audio-recorded per-task answers and the final semi-structured interview.

Within-subjects contrasts use Wilcoxon signed-rank tests with matched-pairs effect size $r$, alongside descriptive $M \pm SD$ for both conditions.
Reasoning-sensitive measures are read from the first block only (six versus six; Mann--Whitney $U$, Cliff's $\delta$), reported descriptively because twelve participants cannot power a between-subjects hypothesis test.
Spearman correlations relate background ratings to perceived and behavioral outcomes. 
Interview and per-task answers were transcribed, translated, and \jw{deductively} grouped by research question\jw{, first by one author, checked by another author}.
\begin{table}[h!]
\centering
\caption{\jw{Session overview per condition: mean (SD), with the Wilcoxon signed-rank $p$ where the measure is comparable.}}
\label{tab:overview}
\setlength{\tabcolsep}{4pt}
\scalebox{0.7}{%
\begin{tabular}{l r r r}
\toprule
 & \systemName & Baseline & $p$ \\
\midrule
Adaptive versions (rules / static variants) & 4.6 (2.6) & 3.2 (1.5) & --- \\
Session length (min) & 27.6 (6.3) & 16.8 (2.7) & --- \\
Adaptive versions per minute & 0.16 (0.07) & 0.20 (0.10) & .57 \\
Raw NASA-TLX overall (0--10) & 4.7 & 4.7 & 1.0 \\
SUS (0--100) & 67.9 (16.6) & 69.8 (21.7) & .45 \\
\bottomrule
\end{tabular}}
\Description{A table with five rows comparing AdapTest and the baseline. AdapTest sessions produced 4.6 rules on average against 3.2 static variants, ran 27.6 minutes against 16.8, and yielded 0.16 adaptive versions per minute against 0.20, a difference with p of .57. Raw NASA-TLX overall workload was 4.7 in both conditions, and SUS was 67.9 against 69.8 with p of .45.}
\end{table}

\section{Results and Findings}
\begin{figure}[h!]
    \centering
    \includegraphics[width=0.7\linewidth]{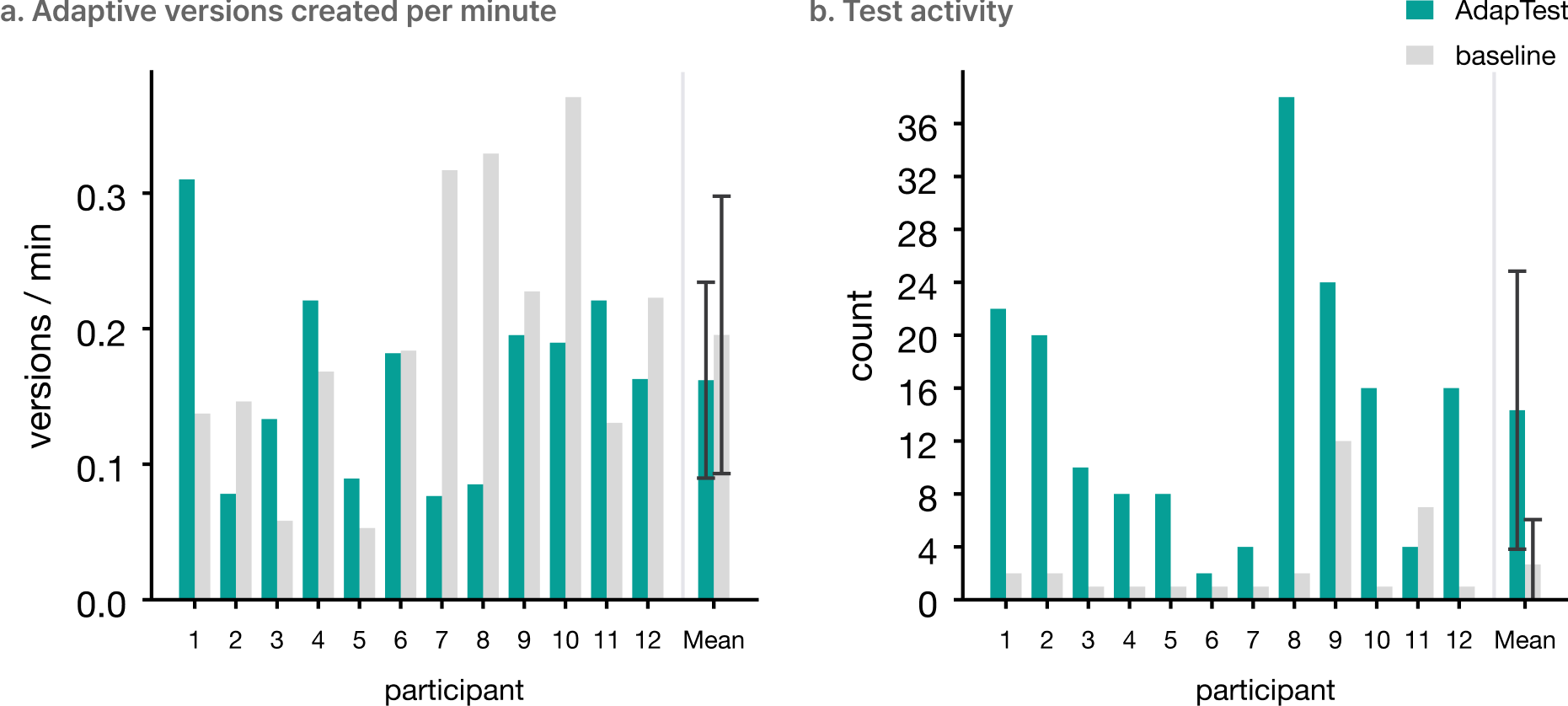}
    \caption{Cross-condition comparison per participant: (a) adaptive versions authored per wall-clock minute, \systemName beside baseline; (b) test activity---\systemName Test Mode runs beside baseline headset-on inspections.}
    \Description{Two grouped bar charts comparing \systemName with the baseline for participants P1 through P12. In panel a, adaptive-version rates vary across participants; the baseline mean is moderately higher than the \systemName mean, although the standard-deviation error bars overlap. In panel b, \systemName produces more test activity for every participant except P11. \systemName Test Mode counts range from 2 to 38, whereas baseline headset inspections are generally 1 or 2, with higher counts of 12 for P9 and 7 for P11. Separated mean bars with plus-or-minus-one-standard-deviation error bars appear at the right of both panels.}
    \label{fig:compare}
\end{figure}

In the smart-home domain, designers authored more adaptive behavior with \systemName than they did in the office domain with their familiar workflow, at comparable workload and usability.
With \systemName, all 12 participants captured how the interface should adapt as 55 rules \sout{($M = 4.6$, $SD = 2.6$, range 2--11)}\jw{(Table~\ref{tab:overview})}, and every rule produced a working adaptive version of the design---a live, condition-specific version, not just a note describing one.
In the baseline, participants produced 38 adaptive versions as static variants \sout{($M = 3.2$, $SD = 1.5$)}, with the conditions that should trigger them only written beside the design or spoken.
Because the two tasks ran on different time budgets by design \sout{($M = 27.6$ minutes, $SD = 6.3$ with \systemName; $M = 16.8$, $SD = 2.7$ in the baseline)}, we read this comparison as an authoring rate (Figure~\ref{fig:compare}a): \sout{participants produced adaptive versions at $M = 0.16$ per minute ($SD = 0.07$) with \systemName and $M = 0.20$ per minute ($SD = 0.10$) in the baseline, a difference that is not reliable}\jw{the two rates did not differ reliably}\footnote{Throughout, we call a difference \emph{reliable} when a statistical test indicates it is unlikely to be due to chance (here, $p < .05$).} \sout{($W = 31$, $p = .57$)}.
The volume difference thus reflects sustained authoring over a longer session at a comparable rate.
Self-reported workload was comparable across conditions: raw NASA-TLX showed no reliable difference in overall workload \sout{($M = 4.7$ vs.\ $4.7$ of 10, $p = 1.0$)}. Performance was the only subscale approaching a reliable difference, with participants reporting marginally greater task success with \systemName ($M = 7.2$ vs.\ $5.8$ of 10, $p = .059$, $r = .39$; all other subscales $p \geq .23$).
SUS scores also did not differ reliably between \systemName \sout{($M = 67.9$, $SD = 16.6$)} and the baseline \sout{($M = 69.8$, $SD = 21.7$; $W = 24$, $p = .45$)}.
We report behavioral findings and quantitative results for each research question, followed by \sout{usage-context feedback}\jw{participants' preferences between the two workflows}.

\subsection[Writing Adaptive Behavior on Specific Elements]{\sout{RQ1: }Writing Adaptive Behavior \sout{as Rules }on Specific Elements}

\sout{\systemName moved the description of}\jw{With \systemName, participants recorded} how the interface should change \sout{from side notes to}\jw{as} working rules attached to specific elements\jw{. In the baseline, the same intent stayed in notes and speech}.
Of the 55 authored rules, 44 bound specific elements from the original design (2 to 9 elements per rule, $M = 5.6$, $SD = 1.8$) and 11 applied to a whole screen.
Participants rated both mechanisms as helpful for specification: selecting target elements helped them state \emph{what} should change ($M = 6.4$ of 7, $SD = 0.8$), while writing IF/THEN rules helped them state \emph{when} it should change ($M = 6.2$, $SD = 1.1$).
P1 contrasted this with their usual practice: \textit{``Setting the IF/THEN was much easier; in Figma you could really only write down in text what should happen in which situation, so specifying it here was far easier.''}

In the baseline artifacts, activation logic was written as text or verbally explained.
We coded the twelve baseline Figma files alongside the session recordings: seven participants recorded triggering conditions as canvas annotations or layer names, while the other five explained them only during the walkthrough---P8's two-stage proximity fade, for example, existed as four overlapped static states whose triggering logic was described in speech alone.
Thresholds stayed symbolic where \systemName participants set numbers: P12 annotated the baseline variants with literal placeholders (\textit{``distance $>$ N m''}), explaining that only testing could determine the value.
Baseline participants also reached beyond the staged factors: six of twelve built adaptive versions triggered by time, touch, eye gaze, hand-reach, or the user's own movement---trigger types their static tool could neither encode nor fire.
Consistent with this gap between intent and artifact, baseline participants rated \emph{``I could clearly communicate when the interface should change using this workflow''} at $M = 4.4$ of 7 ($SD = 1.9$), compared with 6.2 for the corresponding \systemName rule item.

Rules could also be reused instead of remade as separate copies.
Six of twelve participants carried at least one rule condition onto a second or third screen (13 replicated conditions in the final projects, and P9 applied one distance rule across all three screens).
P1, whose five replicated conditions span two screens, described the mechanism: \textit{``the other designs follow the same principle---reduce detail when far---so after setting it once I could just paste the rules over to them.''}
Authoring inside the rules still had limits.
P11 found detailed content work harder in the headset (\textit{``With Figma, authoring the window content itself was much easier---adding, editing, changing text''}), but distinguished it from adaptive authoring, describing Figma output as \textit{``making a pretty poster and sticking it up''} whereas \systemName \textit{``makes you consider \ldots designing the interface around actions and spatial relations.''}

\subsection[Noticing and Covering Overlapping Conditions]{\sout{RQ2: }Noticing and Covering Overlapping Conditions}

\begin{figure}[t!]
\begin{minipage}[c]{0.42\linewidth}
    \centering
    \includegraphics[width=\linewidth]{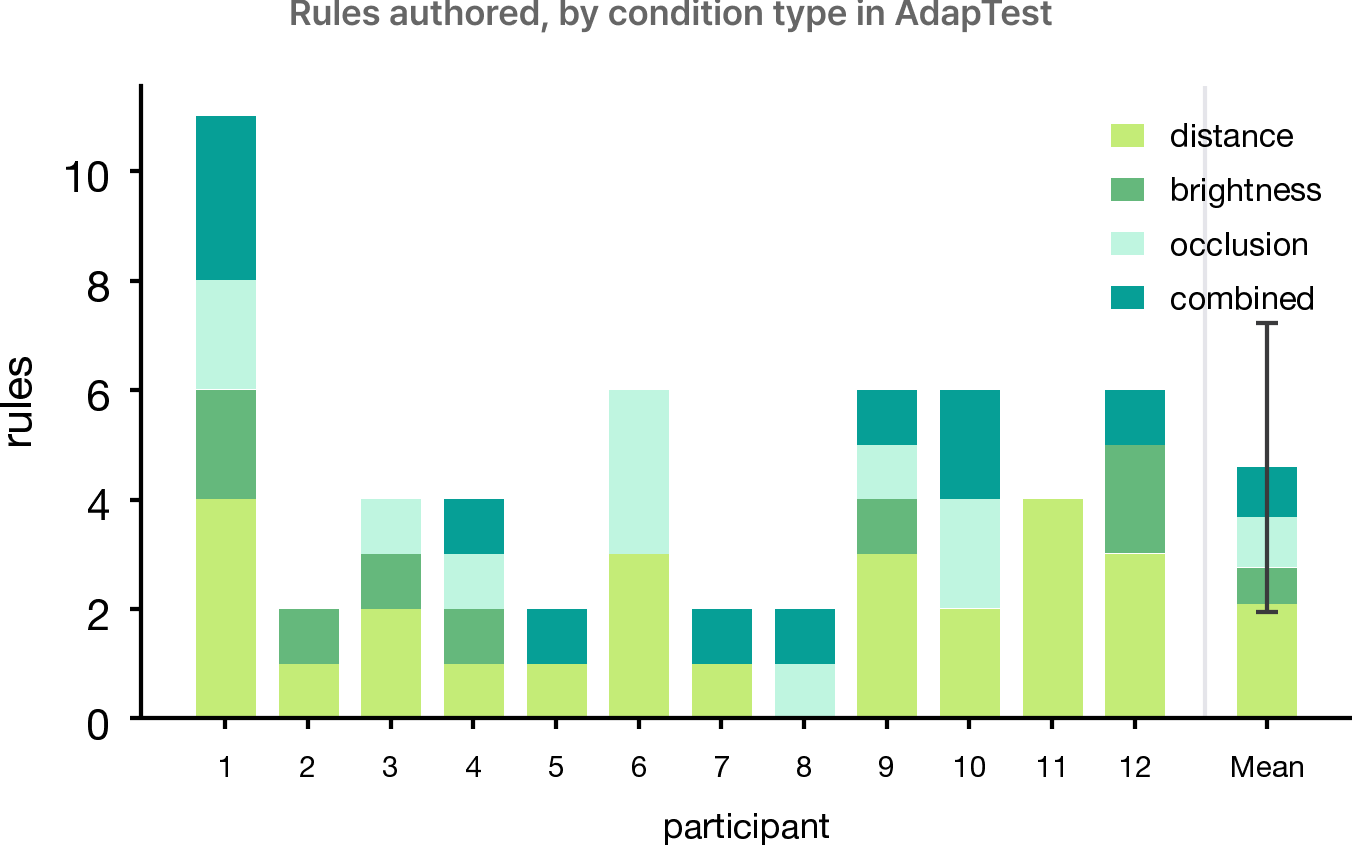}
\end{minipage}\hfill
\begin{minipage}[c]{0.55\linewidth}
    \centering
    \setlength{\tabcolsep}{4pt}
    \scalebox{0.7}{%
\begin{tabular}{l r r r}
\toprule
Condition & Rules & Participants & Median threshold \\
\midrule
distance $>$ $x$ & 16 & 8 & 2.9\,m \\
distance $<$ $x$ & 8 & 4 & 2.55\,m \\
brightness $<$ $x$ & 7 & 5 & 0.36 \\
brightness $>$ $x$ & 1 & 1 & 0.80 \\
occlusion $>$ $x$ & 6 & 4 & 0.66 \\
occlusion $<$ $x$ & 5 & 3 & 0.00 \\
brightness + distance & 8 & 5 & \\
distance + occlusion & 3 & 3 & \\
distance range & 1 & 1 & 1--2\,m \\
\midrule
Total & 55 & 12 & \\
\midrule
Styling output & Rules & Participants & \\
\midrule
hide & 39 & 12 & \\
opacity & 24 & 11 & \\
color & 15 & 8 & \\
size & 14 & 8 & \\
reorder & 6 & 5 & \\
\bottomrule
\end{tabular}}
\end{minipage}
\par\smallskip
\begin{minipage}[t]{0.42\linewidth}
    \captionsetup{skip=0pt}
    \caption{Rules per participant by condition type. The right bar is the mean with $\pm$1 SD.}
    \Description{A stacked bar chart showing final-project rule counts for participants P1 through P12. Total counts range from 2 to 11 and are divided into distance-only, brightness-only, occlusion-only, and combined conditions. All twelve participants addressed distance either through a distance-only or combined rule; eleven have a distance-only segment. Eight participants authored at least one combined rule. P1 authored the most rules, with 11; all other participants authored between 2 and 6. A separated bar at the right shows a mean total of 4.6 rules, with an error bar spanning plus or minus one standard deviation.}
    \label{fig:rules}
\end{minipage}\hfill
\begin{minipage}[t]{0.55\linewidth}
    \captionsetup{skip=0pt}
    \captionof{table}{\jw{Authored rules by condition, and with styling outputs. A rule may use several outputs, and six rules used none.}}
    \Description{Two small tables. The first lists the 55 final rules by condition: distance greater than a threshold in 16 rules by 8 participants with a median of 2.9 meters, distance less than in 8 rules by 4 participants with a median of 2.55 meters, brightness less than in 7 rules by 5 participants with a median of 0.36, brightness greater than in 1 rule, occlusion greater than in 6 rules by 4 participants with a median of 0.66, occlusion less than in 5 rules by 3 participants with a median of 0.00, brightness combined with distance in 8 rules by 5 participants, distance combined with occlusion in 3 rules by 3 participants, and one distance-range rule between 1 and 2 meters; the total is 55 rules by 12 participants. The second lists styling outputs: hide in 39 rules by all 12 participants, opacity in 24 by 11, color in 15 by 8, size in 14 by 8, and reorder in 6 by 5.}
    \label{tab:rulepatterns}
\end{minipage}
\end{figure}

Participants covered more of the staged environmental factors when authoring with rules.
With \systemName, a factor counts as covered when an authored rule senses it; in the baseline, a factor counts even when its condition exists only in the participant's spoken explanation, a looser standard\jw{, so the comparison is descriptive}.
Even under this looser baseline criterion, participants addressed more staged factors with \systemName ($M = 2.4$, $SD = 0.7$; six of twelve covered all three) than in the baseline ($M = 1.4$, $SD = 1.2$; three of twelve covered all three). This within-subjects difference was reliable (Wilcoxon $W = 3.5$, $p = .027$, $r = .64$). \jw{The comparison also differs in task domain and time budget, and} \sout{Because} coverage could carry across blocks, \jw{so} we \jw{read it as a workflow difference and} interpret it alongside the more conservative first-block comparison below.
The authored rules addressed all three sensed conditions (Figure~\ref{fig:rules}): distance in 36 rules ($N = 12$), brightness in 16 ($N = 9$), and occlusion in 14 ($N = 8$). Combined rules were counted under both constituent factors; consequently, these counts overlap and sum to more than 55.
Even with \systemName, coverage was not complete: every participant covered distance, but three still skipped brightness and four skipped occlusion, and P7 even covered occlusion in the baseline yet not with \systemName. Rule authoring increased coverage but did not ensure that every factor was addressed.
\jw{Table~\ref{tab:rulepatterns} summarizes the patterns across the 55 final rules. Hiding elements at a far distance was the modal rule, in 15 of the 16 single distance-greater-than rules, and compound rules mostly paired low brightness with a distance condition. Six rules carried a condition but no styling change.}

Combined conditions were particularly more common with \systemName. 
Nine of twelve participants authored at least one combined rule with \systemName (twelve such rules in all, eleven surviving in the final projects), against two of twelve who built a combined variant in the baseline. 
P4 named the difference: \textit{``In Figma you can't have two conditions at once, so you keep two separate variants and make a whole new variant for when both apply.''}
The same direction appears in each participant's first task, before either tool's practice could carry into the other block: those who started with \systemName covered $M = 2.5$ factors, and those who started with the baseline covered $M = 1.0$ ($U = 30.0$, $p = .054$, Cliff's $\delta = 0.67$). Both groups received the same brief, which named no environmental factors. The observed difference is therefore consistent with \systemName making coverage gaps visible during authoring, whereas the baseline provided no equivalent cue. With six participants per group, we read it as a supporting pattern.

The rule-attached previews---live renderings of how each rule changes the design, generated and updated by the tool---were the capability participants endorsed most strongly, and they helped designers spot problems as they worked.
The preview item scored highest of the five ($M = 6.9$ of 7, $SD = 0.3$), and participants described editing from the previews directly---P4: \textit{``More than the test, I mostly looked at the preview next to the rule and made changes based on that,''} and P7: \textit{``because hides showed up immediately, I tried hiding one more element than planned, saw it looked fine, and kept it.''}
Baseline use further highlighted the value of side-by-side comparison.
In the baseline block, P7 \textit{``regretted not bringing the other version along to compare''} when checking far and near placements.

Suggestions received the lowest rating among the five capabilities ($M = 5.3$ of 7, $SD = 0.8$), but still contributed to the authored rule sets.
Four participants accepted seven suggestions in total, and the five accepted combination suggestions account for five of the twelve combined rules above.
P12 was shown seven suggestions and accepted none, yet covered all three staged factors with self-authored rules, and still projected the value forward: \textit{``the suggestion feature could catch cases I was missing---when things get complicated, it could really help.''}
P11 recalled a suggestion catching a combined case: \textit{``the house could be dark \ldots when the rule suggestion popped up during the pre-test earlier, it felt really convenient.''}

\begin{figure*}[h!]
    \centering
    \includegraphics[width=\linewidth]{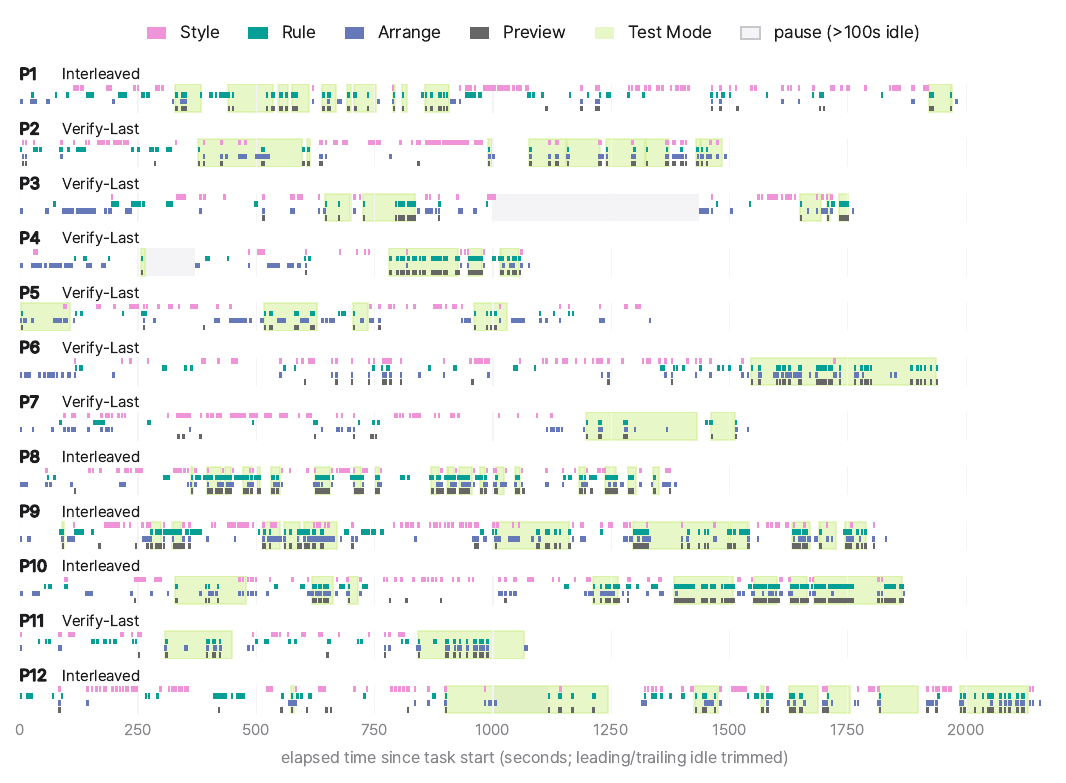}
    \caption{Workflow timelines of all twelve \systemName sessions: activity lanes (Style, Rule, Arrange, Preview) over elapsed time, with Test Mode spans as full-height bands and idle pauses hatched.}
    \Description{Twelve horizontal timelines, one per participant, show Style, Rule, Arrange, and Preview activity over as much as approximately 35 minutes. Pink marks indicate style activity, teal marks rule activity, blue marks arrangement, gray marks previews, light-green full-height bands indicate Test Mode, and hatched gaps indicate pauses longer than 100 seconds. P1, P8, P9, P10, and P12 are labeled Interleaved, with Test Mode recurring across their sessions. The other seven participants are labeled Verify-Last, with testing generally concentrated in later portions of their workflows.}
    \label{fig:timelines}
\end{figure*}

\subsection[Testing Rules in the Real Environment]{\sout{RQ3: }Testing Rules in the Real Environment}

Participants \sout{interleaved authoring with}\jw{ran} repeated \sout{testing throughout}\jw{tests during} the session (Figure~\ref{fig:timelines}).
They ran Test Mode $M = 14.3$ times per session ($SD = 10.5$, range 2--38), completed $M = 10.2$ test--refine cycles ($SD = 4.6$), and edited rule conditions and thresholds $M = 46.1$ times ($SD = 27.3$).
Testing occupied about a quarter of active session time ($M = 26\%$, $SD = 9\%$).
Test Mode hides the editing panels while it runs, so a participant who spots a problem during a test must stop the test before making the fix.
We therefore counted an edit as test-prompted when it occurred during a test or within 2 minutes after a test ended: 430 of the 1{,}218 style and condition edits (35\%) were test-prompted.
Based on build–test alternation, five participants interleaved testing throughout the session, each making at least 15 build–test switches, while seven concentrated testing in later verification passes.

\begin{figure*}[h!]
    \centering
    \includegraphics[width=0.6\linewidth]{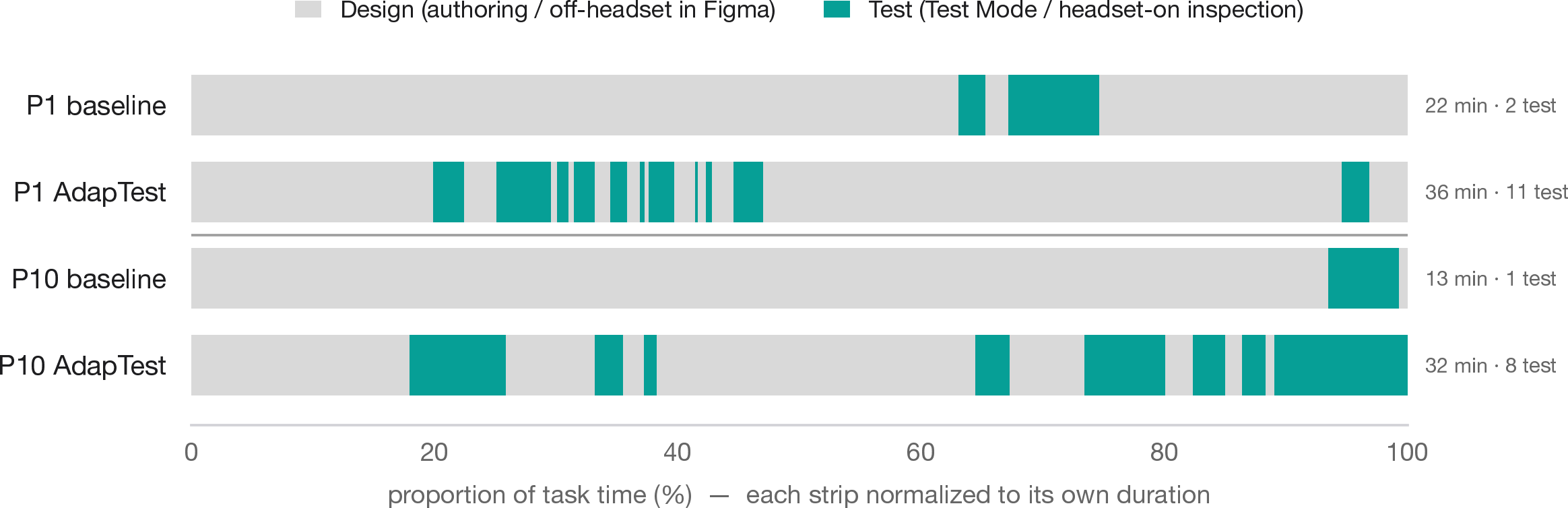}
    \caption{\jw{When testing occurred for P1 and P10 in each condition (strip normalized to own session length).}}
    \Description{Four horizontal strips, one per participant and condition, with teal spans marking headset-on testing on a 0 to 100 percent time axis. In both baseline strips testing is a single late block; in both AdapTest strips short testing spans recur across the session. Session length and test count are listed beside each strip.}
    \label{fig:compare_timeline}
\end{figure*}

The baseline afforded far less of this in-context checking.
Participants could run \systemName Test Mode immediately during authoring ($M = 14.3$ runs). In the baseline, testing required donning the headset to inspect exported images and occurred much less frequently ($M = 2.7$ inspections). This difference in the frequency of in-context checking \sout{was reliable (Wilcoxon $W = 2.5$, $p = .002$, $r = .89$;}\jw{is descriptive (}Figure~\ref{fig:compare}b), \sout{although}\jw{because} the two measures represent different activities: running a live Test Mode session and donning the headset to inspect a static image\jw{, and the baseline could not run a live test by construction}.
The two workflows also placed testing at different times (Figure~\ref{fig:compare_timeline}): with \systemName, testing wove through authoring, whereas the baseline separated a long stretch of off-headset Figma design from a few late inspections.

\begin{figure*}[h!]
    \centering
    \includegraphics[width=1\linewidth]{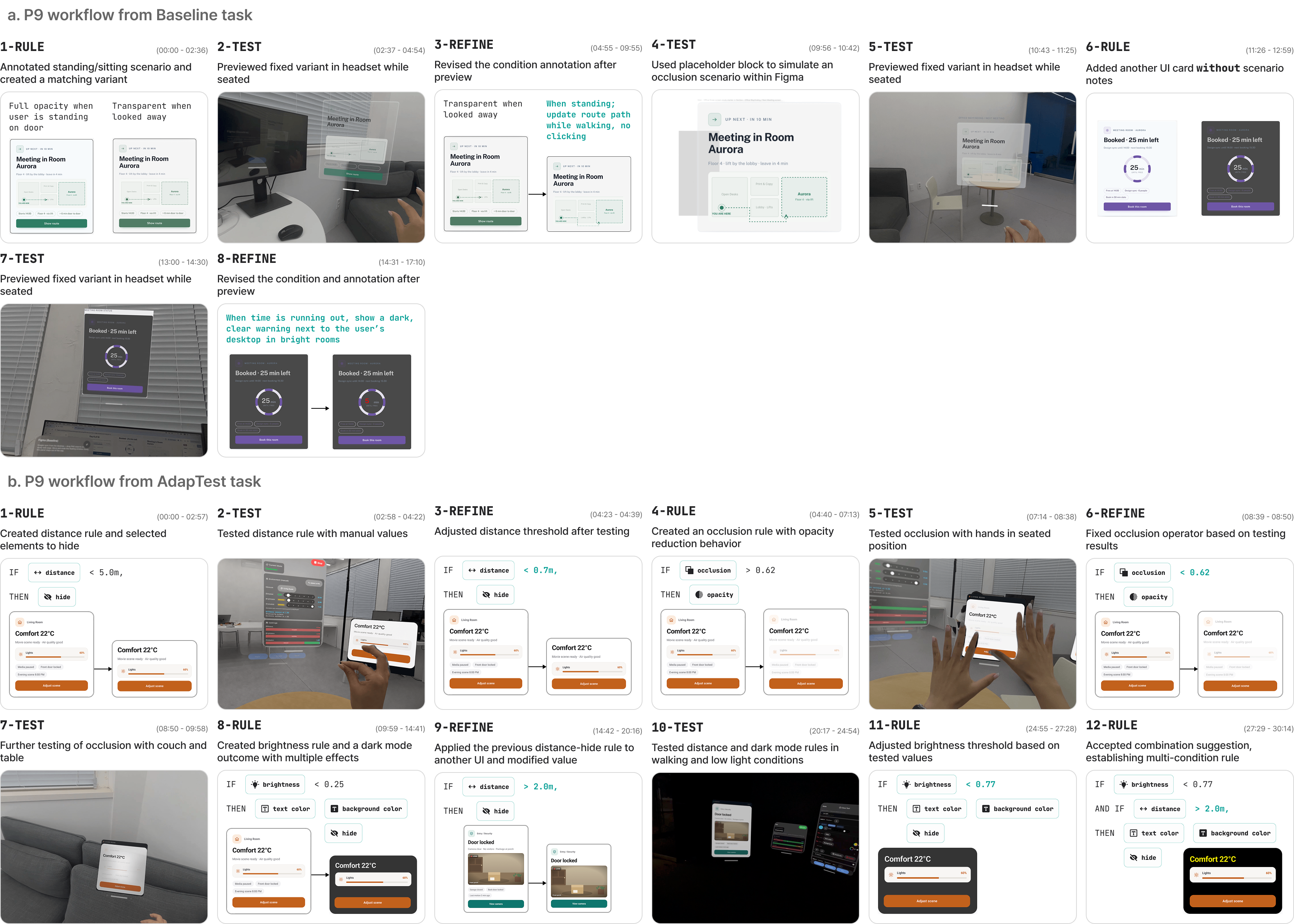}
    \caption{P9's workflow (completed the baseline task first), split into numbered steps and labeled by phase: \textbf{a)} baseline condition, \textbf{b)} \systemName condition.}
    \Description{A two-row storyboard of participant P9's workflow in the two study conditions; each numbered step is a panel with a phase label, a timestamp, and a screenshot or headset photo. Row (a), the baseline task, has eight steps over about seventeen minutes: P9 annotates a standing/sitting scenario and makes a matching variant, previews the fixed variant in the headset while seated, revises the condition annotation, uses a placeholder block in Figma to simulate occlusion, previews again, adds another card without scenario notes, previews it, and revises its condition and annotation. These panels show Figma cards with written condition notes and photographs of static cards floating in the room. Row (b), the \systemName task, has twelve steps over about thirty minutes that alternate Rule, Test, and Refine: P9 creates a distance rule to hide elements (IF distance less than 5.0m THEN hide), tests it with manual values, lowers the threshold to 0.7m, creates an occlusion rule that reduces opacity (IF occlusion greater than 0.62), tests it with hands while seated, corrects the operator to less-than after testing, tests further with a couch and table, creates a brightness dark-mode rule with several effects (IF brightness less than 0.25 THEN text color, background color, and hide), reuses the distance-hide rule on another screen with a new value (distance greater than 2.0m), tests distance and dark mode while walking in low light, adjusts the brightness threshold to 0.77 from tested values, and accepts a combination suggestion that merges brightness and distance into one multi-condition rule. These panels show IF/THEN rule cards with before-and-after previews and photographs of the interface responding in the headset.}
    \label{fig:p9}
\end{figure*}

Participants rated Test Mode highly for checking whether adaptations worked in XR ($M = 6.2$ of 7, $SD = 1.2$), compared with $M = 4.3$ ($SD = 2.0$) for the corresponding baseline item.
P11 mapped the loop onto a familiar practice: \textit{``It was very similar to working in Unity---fix a bit, hit play, test it, then revise again.''}
\sout{Testing converted}\jw{Participants described setting} thresholds and styling from guesses into observed values \jw{during tests}.
P9 calibrated a distance rule against the live reading (Figure~\ref{fig:p9}): \textit{``Watching the live reading, I measured how far the threshold should be---it was about 2m to that door,''} and P3 described the same conversion: \textit{``only by actually trying did I learn at what closeness or distance it applies.''}
Styling changed on observation too---P12: \textit{``when I actually tested it I thought `this is brighter than I expected, I should reduce it' and redid it,''} and P8 dropped the ``Living Room'' label from a reduced state after testing showed it unnecessary.
P1 contrasted this with the baseline's stand-ins---\textit{``for `dark' you'd just fill a black rectangle behind it, so having it actually happen was much better''}---and P10 located the difference: \textit{``with Figma the work happens only in my imagination.''}

The physical room bounded what Test Mode could enact, and participants split on what should stay physical.
P1 wanted a farther condition than the room allowed---\textit{``I wanted to see how it looks when it's tiny from the far end, but the test couldn't show me that far''}---and P2 wanted more extreme conditions than the session offered.
P8 asked for a virtual distance control for this reason, yet kept one condition physical: \textit{``Occlusion should be tested by physically doing it.''}
P11 explained why embodiment mattered for numbers at all: while authoring, \textit{``I'm sitting still and can't gauge what 1m or 2m means anyway''}---the values became judgeable only in motion.

\subsection[Preferences Between Workflows]{\sout{Usage-Context Feedback} \jw{Preferences Between Workflows}}

Participants rated all five \systemName capabilities above the midpoint (previews, element selection, rules, Test Mode, and suggestions, in order) and the interviews clarified \sout{how the workflow might fit existing practice}\jw{why they preferred one workflow over the other for a given kind of work. Four reasons recurred: immediacy, embodied judgment of values, precision of content edits, and learning cost}.

Participants placed the workflow inside practices they already knew.
P5 recognized their own habits: \textit{``Using the Vision Pro actually feels much more natural to me, because I usually design by quickly roughing something out, placing it, and looking at it,''} and P9 saw a round trip disappear: \textit{``Designing VR content on a computer means you still have to re-verify it in VR anyway, so with a good system doing it in VR is much better.''}
\sout{The cost participants named was the entry step: they}\jw{Learning cost and content precision favored the baseline. Participants} rated \systemName higher on ``I needed to learn a lot of things before I could get going'' ($M = 3.0$ vs.\ $2.1$ of 5, $p = .016$, $r = .49$), and precise content edits such as changing text stayed easier on the desktop.
Looking beyond the study task, P7 identified where coverage work becomes particularly important in practice: \textit{``For shipping products, companies imagine many very specific edge cases, so this environmental control would be even more useful there,''} and P5 noted the enabling condition: \textit{``normally we never considered [distance and brightness] precisely because everyday sensors can't give exact distance; with accurate data the UI experience could get much better.''}

\section{Discussion}

Our formative study surfaced three findings: designers already adapt concrete interface elements with a small visual vocabulary, they lose track of the growing set of environmental cases, and they can judge thresholds and transitions only in the space itself.
\systemName translated each finding into a design goal: element-targeted rule authoring (DG1), multi-condition preview comparison (DG2), and situated testing of the active rule (DG3).
The comparative study then showed designers taking up all three capabilities while keeping their own workflows: all twelve externalized adaptive intent as working rules, \sout{factor coverage widened}\jw{they covered more of the staged factors}, and testing wove through authoring in two distinct rhythms, interleaved throughout or concentrated in later verification passes.
We discuss what these results mean for each design goal and close with limitations and future work.

\subsection{Authoring Adaptivity as Designer-Owned Rules}

Element-targeted rules kept the condition-appearance mapping in the designer's hands, and designers used that ownership.
All twelve participants stated when and how their interfaces should change as working rules, where the same participants' baseline artifacts held that intent in annotations and speech.
They also rated their ability to communicate change conditions higher with rules (6.2 vs.\ 4.4 of 7).
Prior work has advanced the complementary side of this division: automated systems decide adaptations at runtime through solvers and optimizers \cite{AUIT, semanticAdapt, situationAdapt}\jw{, increasingly with user preferences steering the objectives \cite{ParetoAdapt_2023, AutoOpt_2026}}, and even when designers pre-author the variants, the runtime owns the mapping from context to variant \cite{ContextAdapt}.
Trigger-action tools established that non-programmers can author rich behavior over discrete events \cite{IFTTT, XRSpotlight, ProGesAR, ProInterAR}.
Our study adds that visual designers, given concrete elements and continuous environmental triggers, author this mapping fluently and treat it as part of the design itself---P11 described the shift as designing ``around actions and spatial relations'' where static tools produce posters.

The triggers designers reached for on their own show where rule authoring should extend next.
%wall clock like interior blended design, around their working desk environment or attach to the wall near the door to act as a videodoorphone, similarly also observed from formative study
Six of twelve baseline participants authored states for triggers the study did not stage---time, touch, eye gaze \cite{GazeandSpeech}, hand-reach, and the user's own movement---even though the task brief and staging pointed only at environmental factors.
\jw{These triggers fall into four families beyond the environmental one \systemName senses: interaction (touch, gaze, hand-reach), time of day, the user's own movement and proximity, and, in the fixture ideas below, the presence of other people. Each family maps to a different use case, from a museum panel that reacts to a visitor's approach to a home panel that goes quiet at night.}
Several of these families have long precedents: proxemic interaction research established distance, orientation, and movement as interaction dimensions \cite{Marquardt_2011, Ledo_2015, Vermeulen_2015}, and gradual engagement described the staged reveal-on-approach behavior that P8 rebuilt by hand from static states \cite{Marquardt_2012}.
The sensing capabilities these triggers require (e.g., eye tracking, hand tracking, and position tracking) are already available on the headset, and adaptation drivers beyond the environment, such as privacy, are emerging in parallel \cite{Rajaram_2025}.
Extending element-targeted rules to these trigger families would let a single IF/THEN construct carry both interaction- and context-driven adaptivity, keeping the rule set as a single inspectable record of both.

Designers also imagined interfaces that take on a role in the room, not only a position within it.
In both studies, participants pictured panels as interior fixtures (e.g., a glanceable clock blended into a desk, or a panel by the door that becomes a video-doorphone as someone approaches) so the interface's anchor and the activity around it, alongside ambient conditions, shaped how it should behave.
This extends the reveal-on-approach of gradual engagement~\cite{Marquardt_2012} to authored 2D interfaces: treating an interface's placement and the presence of people near it as conditions, and its role---calm and glanceable at rest, interactive on approach---as an output, would let \systemName author interfaces that behave as ambient objects anchored in a space.

\subsection{Seeing Coverage Before Deployment}

\jw{With} \sout{Condition-specific}\jw{condition-specific} previews\jw{, coverage became an inspection task, and in the baseline it stayed a memory task} \sout{turned coverage from a memory task into an inspection task}.
Participants addressed more of the staged factors with the tool (2.4 vs.\ 1.4 of 3), treated combined conditions as routine (nine of twelve authored combined rules), and rated previews highest of the five capabilities (6.9 of 7)---editing directly from them.
Comparing interface alternatives side by side is an established aid to design judgment \cite{Dow_2011, Hartmann_2008, Swearngin_2020, AB4Web}.
Those comparisons weigh alternatives a designer freely explores, and condition-indexed previews adapt the same aid to a different question: whether one design stays legible and usable under each situation it will meet---a concern readability research in AR documents across lighting, background, and distance \cite{Cauz2024}.
Coverage still did not become complete---brightness and occlusion trailed distance, and one participant covered occlusion only in the baseline---so visible previews help designers inspect coverage without ensuring they act on every gap.
The shift from remembering conditions to inspecting them is not specific to XR: any design that must hold across a space of conditions---responsive layouts across screen sizes, light and dark themes, or accessibility variants---meets the same risk of uncovered cases, and condition-indexed previews with coverage strips \sout{carry}\jw{could carry} the same inspection aid to those settings\jw{, as keyframe authoring already spans screen size, ability, and distance for 2D interfaces \cite{FrameKit_2024}}.

The suggestion results sketch how assistance might close those remaining gaps.
Accepted combination suggestions seeded five of the twelve combined rules, so acceptance did direct work.
At the same time, P12 saw seven suggestions, accepted none, and covered all three factors with self-authored rules, while rating the feature as able to ``catch cases I was missing.''
Suggestions may therefore help twice: as rules a designer adopts, and as prompts that provoke the designer to author a response of their own.
Assistance for coverage could lean into the second mode---surfacing an uncovered condition as a visible gap beside the previews and leaving the authored answer to the designer.

\subsection{Testing as Part of Authoring}

Situated testing became a design activity woven through authoring\jw{ in the \systemName condition}, where XR practice studies had described it as a costly separate step \cite{Ashtari_2020, CoCreatAR}.
Testing filled about a quarter of active session time; 35\% of edits followed a test within 2 minutes; and \jw{participants described setting} thresholds \sout{became}\jw{from} observed values---P9 set a distance threshold by walking to a door and reading 2m.
Authoring rate \sout{matched}\jw{did not differ reliably from} the baseline while sessions ran longer, so the added time went into testing and refining.
The rhythms differed and the tool supported each: seven participants concentrated testing in later verification passes, while five interleaved tests from the start---at cadences from rapid edit--test alternation (P8 ran 38 tests around two rules) to slower build--test cycles.
In-situ authoring tools established creation in the target environment \cite{DART, PintAR, Rapido, ShapesXR, Iyer_2025}, where testing reflects the conditions the room offers at that moment.
The active-rule Test Mode adds the ability to dial a condition on demand and know which authored state is showing.

\jw{Seen across workflow types, the trade-off runs along where checking happens.
A fully 2D workflow, as in the baseline, keeps content edits precise and reuses the design system, and it lets designers name any trigger they can imagine. Each condition, though, stays as text or speech, and every check needs a headset round trip that P5 and P9 described as a separate step (Section~6.4).
A fully XR workflow, as in ShapesXR \cite{ShapesXR} or the Unity play loop P11 compared Test Mode to, gives immediacy and spatial judgment from the start. It gives up the fine content editing and asset reuse that participants missed in the headset. We did not evaluate such a tool, so this reading rests on prior work and on participants' comparisons.
\systemName sits between the two. The 2D design stays the source, rules execute and are tested in place, and coverage is visible in previews, at the entry cost participants reported.
Which arrangement suits a project depends on whether its conditions can be judged from the desk, how much fine editing remains, and how often the design must be checked in the room.}
Participants' requests draw a concrete design space for simulating conditions the room cannot produce.
P1 wanted a viewing distance beyond the room's far wall, P2 wanted more extreme conditions than the session offered, and P8 asked for a virtual distance control---yet P8 insisted occlusion ``should be tested by physically doing it,'' and P11 could judge what 1m meant only while moving.
Simulation and embodiment thus split by condition: distance and lighting invite virtual extension---simulating AR conditions from within VR has precedent \cite{Cavallo_2019}, and headset immersive environments could present a twenty-meter hall or a dark room inside a small lab---while occlusion and threshold judgment draw their value from the body.
Future situated-testing tools could offer simulated ranges for the former while keeping the latter physical, widening testable conditions without leaving the authoring session.
%muliple user for design consistency, easier handoff or collaborating teamwork with other designers or the development team to describe what they want for end user to experience , they could also experience that will shorten the time to iterate possibly?

The same executable, in-place rules suggest a role in collaboration that our single-user sessions did not test.
Because an authored rule set runs and can be entered in Test Mode, a teammate or developer could step into the space and experience the intended adaptive behavior directly, where a mockup or written specification leaves that behavior to the imagination---a shared, situated reference between designers and the developers who implement the result~\cite{Borsting_2022, CoCreatAR, XSpace}.
Shared rule sets and semantic roles could likewise keep a team's adaptive behavior consistent across screens and products, and experiencing the behavior together might shorten the design--development iterations that adaptive intent otherwise invites.

\subsection{Limitations and Future Work}

The main limitation of our study is that each tool was tested on its own fixed task domain, so the comparison is not a pure tool contrast.
We fixed this condition-to-domain pairing to keep the physical staging comparable, and the results therefore compare \systemName-on-home with baseline-on-office under matched task structure.
The shared screen scaffold and the first-block contrast mitigate this, and a crossed replication would isolate the tool effect.
The twelve single-session participants likewise capture only first-contact use, representing one point in the broader space of evaluation strategies \cite{ToolkitEvaluation}, and the staged conditions were bounded by what one study room could produce.
Longer-term use---on designers' own projects, with the learning cost amortized---remains to be studied.
\jw{The evaluation therefore supports claims about the authoring and testing process and about participants' perceptions in staged tasks. It does not show that the resulting interfaces are more usable for end users, that deployed adaptations behave better, or that developer handoff improves, since none of these was measured.}

The tool itself has limitations that point to future work.
Participants rated it higher in terms of needing to learn before starting, and fine content edits remained easier on the desktop.
Its three sensing factors cover part of designers' adaptive intent, and the simulation directions above, together with the additional trigger families in Section 7.1, would broaden both the authorable and testable ranges.
\jw{The sensing itself is uncharacterized. Brightness is normalized to the range seen since launch, so a threshold is relative to that session and to camera exposure. Occlusion in the study came from tracked hands only, so physical obstacles and other virtual windows did not register. None of the three readings was validated against an instrument, so thresholds are comparable within a session and not across rooms.}
Environmental sensing also carries ethical weight: the tool senses the surroundings only while the designer is testing, and the study stored structured interaction events without camera frames, room reconstructions, or bystander recordings, but tools that ship authored rules into everyday spaces will inherit the privacy questions of continuous sensing \cite{Rajaram_2025}\jw\jw{. Other people in the room would be sensed without having agreed to it. A version that reads camera frames would also have to say what it keeps and for how long}.
Finally, the exported rules and variants now travel back to desktop tools, and evaluating that handoff with the developers and collaborating design teams who receive it---where XR development practice \cite{Borsting_2022} and spatially distributed collaboration \cite{XSpace} set their own requirements---would be the next step.
\section{Conclusion}
We presented \systemName, an authoring tool that lets interface designers make their 2D designs adaptive to the environments those designs will inhabit, through element-targeted rules, attached condition previews, and situated testing of the active rule.
Designers pick the elements to change by tapping them on the rendered design, state when and how they should change as IF/THEN rules over distance, lighting, and occlusion, compare each rule's outcome in previews beside the default design, and validate the behavior in place under manual controls or live sensing.
The tool accommodates designers' own working rhythms, supporting testing woven through authoring as well as verification concentrated at the end.
\systemName was designed from a formative study with twelve designers and evaluated with twelve screen-rooted designers against their familiar desktop-and-preview workflow.
Participants took up all three capabilities, externalizing adaptive intent as working rules, \sout{widening coverage of environmental conditions}\jw{covering more of the staged environmental conditions}, and \sout{grounding thresholds in tested observation}\jw{refining thresholds and styling from what they observed in tests}, at workload and usability comparable to the tools they already knew.
Future work will extend rules to the trigger families designers already reach for, simulate conditions a single room cannot stage, and follow authored rules into longer-term practice and handoff.

\begin{acks}
This work was supported by the National Research Foundation of Korea (NRF) grant funded by the Korea government (MSIT)(RS-2024-00337803).
\end{acks}

\bibliographystyle{ACM-Reference-Format}
\bibliography{references}

\appendix

\section{Formative Study: Demographics}
\label{appendix:demographic}
\begin{table}[h!]
    \setlength{\tabcolsep}{3pt}
    \centering
    \resizebox{\columnwidth}{!}
    {\begin{tabular}{c|c|c|c|c|c|c|c|c|c}
        \toprule
        \textbf{ID} & \textbf{Age} & \textbf{Gender} & \textbf{Background} & \textbf{\makecell[c]{XR familiarity\\(1 - 7)}} & \textbf{\makecell[c]{First Time\\ of use}} & \textbf{\makecell[c]{Type of\\ devices}} & \textbf{\makecell[c]{XR Contents\\ Experienced}} & \textbf{\makecell[c]{UI Design\\Experience}} & \textbf{\makecell[c]{XR Design Experience}} \\
        \midrule
        P1 & 25 & F & HCI/UX & 5 & 2020 & Quest & \makecell[c]{Space-related\\ Research} & $4\sim5$ years & XR research prototypes (1 year) \\
        P2 & 24 & F & HCI & 2 & 2022 & Quest & \makecell[c]{Zapworks \&\\ Friend's VR Project} & $3\sim4$ years & None (HCI research only) \\
        P3 & 24 & F & Design & 2 & 2017 & Quest 2 & \makecell[c]{Games \&\\ Entertainment} & $1\sim2$ years & None (usage only) \\
        P4 & 23 & F & \makecell[c]{HCI \&\\ SW Development} & 2 & 2014 & \makecell[c]{Vision Pro \&\\ Unknown} & Game & $1\sim2$ years & None (usage only) \\
        P5 & 24 & M & UX/UI Design & 3 & 2022 & \makecell[c]{iPhone \&\\ Quest} & Game (Pokemon Go) & $3\sim4$ years & Mobile AR interfaces (2 years) \\
        P6 & 24 & F & Industrial Design & 3 & 2023 & Unknown & Game & $2\sim3$ years & None (usage only) \\
        P7 & 24 & F & HCI & 2 & 2023 & Quest & Experiment & $2\sim3$ years & None (research participant only) \\
        P8 & 29 & M & \makecell[c]{Industrial design} & 2 & 2022 & Quest 2 & Game & $<1$ year & None (usage only) \\
        P9 & 24 & F & Industrial Design & 4 & 2023 & Quest & Sports Game & $3\sim4$ years & None (usage only) \\
        P10 & 26 & M & \makecell[c]{HCI \&\\ Industrial Design} & 2 & 2018 & Quest & Experiment & $1\sim2$ years & VR research prototypes (1 year) \\
        P11 & 26 & M &  \makecell[c]{Industrial Design \&\\ Experience Design} & 5 & 2023 & Quest Pro & Game & $4\sim5$ years & None (usage only) \\
        P12 & 25 & M & Robot Interaction & 4 & 2018 & \makecell[c]{Google\\ Cardboard} & Game & $<1$ year & None (usage only) \\
        \bottomrule
    \end{tabular}}
    \caption{Demographic and background information of all formative study participants.}
\end{table}

\section{Formative Study: Interview}
\label{appendix:interview}

The following questions guided the semi-structured interviews:

\begin{enumerate}
    \item Why did you choose to set the number of interfaces to be many/few or complex/simple?
    
    \item Did spatial structures or contexts influence your UI design decisions regarding environmental factors?
    \begin{enumerate}
        \item If the space were larger, more complex, or structured differently, would you have designed the interfaces differently?
    \end{enumerate}
    
    \item What were your reasons for setting the interface size and composition at each viewing location?
    \begin{enumerate}
        \item What was the most important design factor you considered?
    \end{enumerate}
    
    \item You placed \textit{n} interfaces in this space—if you were to redesign, would you change the number of interfaces?
    
    \item Did you consider there to be an optimal window size, either in absolute physical dimensions or relative to the user's field of view?
    
    \item How did you design the UI components (videos, images, text, buttons) to change across different viewing distances?
    \begin{enumerate}
        \item What were your reasons for these component variations?
        \item Did you find certain component types to be particularly effective at specific distances?
    \end{enumerate}
    
    \item For designs targeting the same zone from different viewing positions, why did you make the interfaces similar or different?
    \begin{enumerate}
        \item: How would your interface design have changed if the viewing conditions differed (e.g., distances had been much closer or farther)?
    \end{enumerate}
    
    \item In what aspects did the environment between user and interface most significantly influence your UI design decisions (size, position, content display method, component types, etc.)?
    \begin{enumerate}
        \item If distance did not influence your design, what were the reasons?
    \end{enumerate}
    
    \item What factors determined where interfaces would transition between different design states (distance changes, spatial context changes, etc.)?
    
    \item What aspects of designing adaptive XR interfaces are most challenging?
    
    \item For what types of users or contexts do you think adaptive interface design would be most beneficial?
    
    \item Do you have any final thoughts about the design process or spatial interface design in general?
\end{enumerate}

\section{User Study: Interview Protocol} \label{appendix:study_interview}

We include here the full set of questions used in the user study: 

\subsection*{Post-task walkthrough (baseline block)}
\begin{enumerate}
    \item Please briefly walk us through what you created in this workflow, including anything you expressed only as text notes.
    \item Which situation is this version designed to respond to?
    \item Looking at the artifact alone, how would one know when this version should appear?
    \item What could you check directly in this workflow, and what did you still have to imagine or explain?
\end{enumerate}

\subsection*{Post-task walkthrough (\systemName{} block)}
\begin{enumerate}
    \item Can you walk us through your final design as a whole?
    \item Choose one rule you consider important and explain it: why this target element, this environmental condition, and this change?
    \item Did any part of the design change after you used the previews or Test Mode? What did you observe that led to the revision?
    \item Did a suggested combined rule influence any design decision? If you accepted, edited, rejected, or ignored one, why?
    \item Were there situations you noticed but chose not to address?
\end{enumerate}

\subsection*{Semi-structured interview}
\begin{enumerate}
    \item You used both workflows today. What felt most different between working in Figma and working in \systemName{}?
    \item Which parts made it easier or harder to specify what should change and when it should change?
    \item Compared with your usual design practice, what felt familiar and what felt unfamiliar in \systemName{}?
    \item Was there anything you wanted to express in \systemName{} but found difficult to express?
    \item In what projects or scenarios could you imagine using a tool like this?
    \item Conversely, in what situations would you prefer not to use it?
    \item If you could change one thing about the tool before real use, what would you change? (Probe: at which point of the design--development process would it fit?)
\end{enumerate}

\end{document}